\documentclass[graybox]{svmult}

\usepackage{mathtools}
\usepackage{mathptmx}       % selects Times Roman as basic font
\usepackage{mathrsfs}  
\usepackage{helvet}         % selects Helvetica as sans-serif font
\usepackage{courier}        % selects Courier as typewriter font
\usepackage{type1cm}        % activate if the above 3 fonts are
\usepackage{makeidx}         % allows index generation
\usepackage{graphicx}        % standard LaTeX graphics tool
\usepackage{multicol}        % used for the two-column index
\usepackage[bottom]{footmisc}% places footnotes at page bottom
\usepackage{url}
\usepackage{authblk}
\usepackage[numbers,sort&compress]{natbib}

\usepackage{tabularx}
\newcommand{\AX}[1]{\textnormal{(#1)}}
\newcommand{\scen}{\mathscr{S}}

\newcommand{\Aho}{\mathop{Aho}}
\newcommand{\BMG}{\mathfrak{B}}
\newcommand{\fitch}{\mathfrak{F}}
\newcommand{\LDT}{\mathfrak{L}}
\newcommand{\EDT}{\mathfrak{E}}

\usepackage{xcolor}
\definecolor{gray}{rgb}{0.67, 0.67, 0.67}
\definecolor{bgreen}{rgb}{0.2, 0.7, 0.2}
\definecolor{darkgreen}{RGB}{0, 160, 80}
\definecolor{MHcol}{RGB}{200, 100, 0}

\makeatletter
\def\moverlay{\mathpalette\mov@rlay}
\def\mov@rlay#1#2{\leavevmode\vtop{%
    \baselineskip\z@skip \lineskiplimit-\maxdimen
    \ialign{\hfil$\m@th#1##$\hfil\cr#2\crcr}}}
\newcommand{\charfusion}[3][\mathord]{
  #1{\ifx#1\mathop\vphantom{#2}\fi
    \mathpalette\mov@rlay{#2\cr#3}
  }
  \ifx#1\mathop\expandafter\displaylimits\fi}
\DeclareRobustCommand\bigop[1]{%
  \mathop{\vphantom{\sum}\mathpalette\bigop@{#1}}\slimits@
}
\newcommand{\bigop@}[2]{%
  \vcenter{%
    \sbox\z@{$#1\sum$}%
    \hbox{\resizebox{\ifx#1\displaystyle.9\fi\dimexpr\ht\z@+\dp\z@}{!}{$\m@th#2$}}%
  }%
}
\makeatother
\newcommand{\cupdot}{\charfusion[\mathbin]{\cup}{\cdot}}

\usepackage{amsmath}
\usepackage{amssymb}
\usepackage{wasysym}
\newtheorem{thm}{Theorem}

\newcommand{\lca}{\ensuremath{\operatorname{lca}}}

\newcommand{\ROOT}{\circledcirc}
\newcommand{\LEAF}{\odot}
\newcommand{\SPEC}{\newmoon}

\newcommand{\DUPL}{\square}

\newcommand{\R}{\mathscr{R}}
\newcommand{\F}{\mathscr{F}}
\newcommand{\Striple}{\mathscr{S}}

\makeindex             % used for the subject index
\begin{document}
\setitemindent{(iii)}

\title*{The Theory of Gene Family Histories} 

\author{Marc Hellmuth and Peter F.\ Stadler}
\institute{
  Marc Hellmuth \at Department of Mathematics, Faculty of Science,
	Stockholm University, SE - 106 91 Stockholm,   Sweden \newline 
  Peter F.\ Stadler \at Bioinformatics Group, 
  Department of Computer Science, and
  Interdisciplinary Center for Bioinformatics, University of Leipzig,
  H{\"a}rtelstra{\ss}e 16-18, D-04107, Leipzig, Germany.
  Max Planck Institute for Mathematics in the Sciences,
  Inselstra{\ss}e 22 D-04103 Leipzig, Germany.
  Fraunhofer Institute for Cell Therapy and Immunology,
  Perlickstra{\ss}e 1, D-04103 Leipzig, Germany.
  Institute for Theoretical Chemistry, University of Vienna, %
  W{\"a}hringerstra{\ss}e 17, A-1090 Wien, Austria.  Center for non-coding
  RNA in Technology and Health, University of Copenhagen,
  Gr{\o}nneg{\aa}rdsvej 3, DK-1870 Frederiksberg C, Denmark.  Santa Fe
  Institute, 1399 Hyde Park Rd, Santa Fe, NM 87501, USA\\
\email{marc.hellmuth@math.su.se,studla@bioinf.uni-leipzig.de}
}

\maketitle

\abstract{Most genes are part of larger families of evolutionary related
  genes. The history of gene families typically involves duplications and
  losses of genes as well as horizontal transfers into other organisms. The
  reconstruction of detailed gene family histories, i.e., the precise
  dating of evolutionary events relative to phylogenetic tree of the
  underlying species has remained a challenging topic despite their
  importance as a basis for detailed investigations into adaptation and
  functional evolution of individual members of the gene family. The
  identification of orthologs, moreover, is a particularly important
  subproblem of the more general setting considered here.  In the last few
  years, an extensive body of mathematical results has appeared that
  tightly links orthology, a formal notion of best matches among genes, and
  horizontal gene transfer. The purpose of this chapter is the broadly
  outline some of the key mathematical insights and to discuss their
  implication for practical applications.  In particular, we focus on
  tree-free methods, i.e., methods to infer orthology or horizontal gene
  transfer as well as gene trees, species trees and reconciliations between
  them without using \emph{a priori} knowledge of the underlying trees
  or statistical models for the inference of phylogenetic
    trees. Instead, the initial step aims to extract binary relations among
    genes. }


\
\section{Introduction}

\sloppy

In a typical genome, most genes appear as members of larger families of
\emph{homologous} genes, i.e., genes that share a common ancestor. The
evolutionary history of a gene family involves \emph{speciations}, where a
gene is transmitted to each of the separating lineages, \emph{duplications}
within a genome, \emph{loss}, and sometimes also \emph{horizontal
transfer}, either of individual genes or as a consequence of hybridization
events. These events give rise to \emph{homology relations}: two genes are
\emph{orthologs} (resp., \emph{paralogs}) if their most recent ancestor is
a speciation event (resp., duplication event).  Moreover, \emph{xenologous}
genes are genes that were separated by a horizontal transfer.  Such events
tend to impact gene function. Selective pressures on genes may change due
to different environmental constraints following a speciation, but also
following gene duplications due to subfunctionalization or
neofunctionalization \cite{Lynch:00} of the paralogous copies, and
following gene loss \cite{Cutter:16}. As a consequence, orthologous genes
in closely related genomes \emph{often} have \emph{approximately} the same
function. Paralogs, in contrast, tend to have related, but clearly distinct
functions~\cite{Innan:10,Altenhoff:12,studer2009,GK13}, although exceptions
are not uncommon \cite{nehrt2011}. As a consequence, the accurate
distinction of orthologs and paralogs is a key task for functional genome
annotation. The reliable identification of orthologs also plays a key role
in comparative genomics analyses \cite{Sonnhammer2014}. Moreover,
one-to-one orthologs are the characters of choice in molecular
phylogenetics~\cite{GK13,Ballesteros:16}.

Accurate knowledge of the evolutionary history of a gene family, thus, is
the basis for its comparative analysis and an understanding of the evolution
of its functional portfolio. Gene family histories, however, cannot be
measured directly but have to be inferred from present-day sequence data,
i.e., from measurements of similarities between genes and the genomes in
which they reside. This requires a formal framework in which the
intertwined histories of genes and species can be studied and the
correctness, limits, and accuracy of computational methods can be
assessed. In this chapter we outline the framework of \emph{evolutionary
scenarios}, which comprises phylogenetic trees (or more generally
phylogenetic networks) describing the phylogenies of genes and species
together with a reconciliation maps that embeds the gene phylogeny into the
species phylogeny. We will focus here on the case of tree-like evolution
and only briefly comment the generalization to networks.

\begin{figure}[t]
  \centering
  \includegraphics[width = 0.9\textwidth]{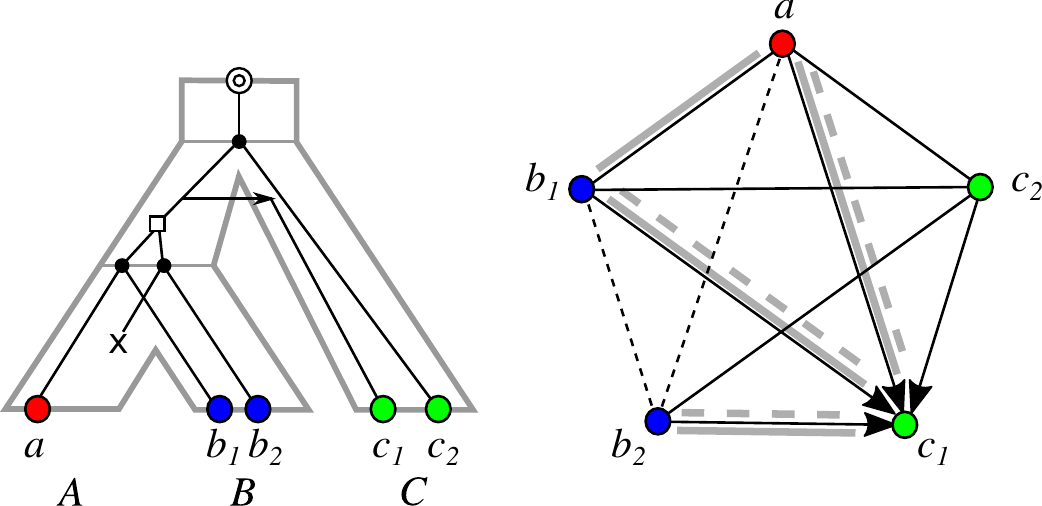}    
  \caption{An evolutionary scenario (left) consists of a species tree $S$
    (tube-like gray outline) into which a gene tree $T$ (black) is embedded
    by means of a reconciliation map that places each vertex of the gene
    tree onto either an edge or a vertex of the species tree (gray
    horizontal lines).  The reconciliation defines the event types on the
    the species tree: \emph{Gene duplication} events ($\square$) are
    located within edges of $S$ and \emph{HGT-edges} (arrow ``$\to$'') in
    which direct offsprings are located in different branches of $S$.
    \emph{Speciations} events ($\bullet$) are located at inner vertices of
    $S$. These have descendants in each branch of the species tree, which
    however may not survive to the present day due to \emph{gene loss}
    ($\boldsymbol{\times}$). Present-day genes are leaves of $T$ that are
    mapped onto leaves of $S$ (shown as leaves at the bottom, colored by
    the species in which they reside, i.e., $\sigma(a)=A$ (red),
    $\sigma(b_1)=\sigma(b_2)=B$ (blue), and $\sigma(c_1)=\sigma(c_2)=C$
    (green).).
    \newline
    The event-labeled gene tree determines the homology relations
    (right). The orthology relation comprises all pairs of genes that are
    connected by a thin solid black edge (without arrows) and correspond to
    pairs of distinctly colored genes whose last common ancestor is a
    speciation event. The paralogy relation comprises all pairs of genes
    that are connected by a thin dashed black edge and correspond to pairs
    of genes whose last common ancestor is a duplication event. The Fitch
    (xenology) relation comprises pairs of genes that are connected by thin
    black arrow-edges.  The later-divergence-time relation comprises pairs
    of genes whose divergence time is below the divergence time of the
    corresponding species (dashed gray lines) and the reciprocal best match
    relation comprises all pairs $(x,y)$ of genes from distinct species $X$
    and $Y$, respectively, for which there is no closer relative $x'\in X$
    for $y$ and $y'\in Y$ for $x$ (solid gray lines).}
  \label{fig:trueHist}
\end{figure}

Evolutionary scenarios define event types that annotate vertices in gene
phylogenies e.g.\ as speciation or duplication and edges as horizontal
transfer. These scenarios in turn form the basis for the formal
definitions for the different homology relations among genes such as
orthology, paralogy, or xenology \cite{Fitch:70,Fitch:00,Darby:17}, see
Fig.~\ref{fig:trueHist} for an illustrative example. These three homology
relations are not the only relations of interest, however.  The
\emph{(reciprocal) best match relation} describes all gene pairs from
distinct species that are evolutionary most closely related
\cite{Geiss:20b,Geiss:20a,Stadler2020,Geiss+2019}. In contrast to orthology
and paralogy, whose definition depends on whether the last common ancestor
was a speciation or gene duplication event, it is possible to estimate best
matches directly from genomic sequence data. Similarly, the
\emph{later-divergence-time (LDT) relation} comprising gene-pairs that have
diverged only after the divergence of the two species in which the genes
reside \cite{Novichkov:04,Schaller:21f,SHL+23} is informative about the
xenology relation, which records whether or not two genes are separated by
a HGT event in their evolutionary history. While xenology cannot be
measured directed, the LDT relation is accessible from practical data
analysis.

The problem of reconstructing a gene family history can be approached in
different ways. \emph{Tree-based} methods start with inferring the gene
tree and the species tree separately using well-established phylogenetic
methods. This leaves the problem of computing the reconciliation map as a
separate optimization problem, typically minimizing the number of loss and
HGT events.  A problematic issue with this approach is that it crucially
depends on accurate tree reconstructions. However, it is difficult to
obtain reliable gene trees in particular for gene families with complicated
histories \cite{Doyle:15}. One remedy is to jointly infer gene trees and
species trees, see e.g.\ \cite{BSD+13,Szollosi:15}.  An alternative
approach are \emph{tree-free} approaches, that estimate binary relations
such as reciprocal best matches directly from sequence similarity data and
then further analyze these relations to extract orthology and paralogy
relations without explicitly constructing trees. It appears that there is
no fundamental difference in the accuracy of present-day tree-based and
tree-free approaches \cite{Altenhoff:09,Altenhoff:19}. 
 Methods to infer
HGT, reviewed in \cite{Ravenhall:15}, fall into three major categories: (1)
tree-based approaches that compute an optimal reconciliation w.r.t.\ some
cost function \cite{Tofigh:11,Chen:12,Ma:18}, parametric methods that use
genomic signatures, i.e., sequence features specific to a (group of)
species identify horizontally inserted material
\cite{Dufraigne:05,Becq:10}, and so-called implicit or indirect methods use
distances between genes that are very small or very large compared to the
evolutionary distances of the respective species as indicators of HGT
\cite{Novichkov:04,Kanhere:09}. In fact, these methods in essence estimate
the LDT relation \cite{Schaller:21f}.

The purpose of this chapter is to give an introduction into the formal
framework of evolutionary scenarios and to review some key mathematical
results describing the relationships between binary relations that can be
measured from similarity data and the homology relation that are of
interest in evolutionary biology. Moreover, we will be concerned with the
question how much, and which, information on the gene family history is
actually encoded in these binary relations.  We shall see that robust
estimates of these binary relations already put tight constraints on gene
trees, species trees, and reconciliations.

\section{Scenarios}

\subsection{Notation}

\par\noindent\textbf{\emph{Graphs}} $G$ are tuples $(V,E)$ consisting of a
vertex set $V(G)\coloneqq V$ and edge set $E(G)\coloneqq E$.  Graphs might
be undirected (edges are 2-elementary subsets of $V$) or directed (edges
are subsets of $V\times V$). All graphs considered in this contribution are
loop-free, i.e., if $\{x,y\}$ or $(x,y)$ is an edge, then $x\ne y$.  For a
directed graph $G$ we write $G_{sym}$ for its underlying undirected graph
with $V(G_{sym})=V(G)$.  In general, we use the term ``graph'' for
undirected graphs and, otherwise, explicitly write ``directed graph''.
Moreover, we use the simplified notation $xy$ for edges $\{x,y\}$ in
undirected graphs or for the case that both $(x,y)$ and $(y,x)$ are edges
in a directed graph. Identifying undirected graphs with symmetric
  directed graphs, furthermore, allows us to make use of subgraph
  relationships between directed and undirected graphs.
We extensively make use of graphs $(G,\sigma)$
equipped with a vertex-coloring $\sigma$. We say that $(G,\sigma)$ is
\emph{properly colored} if $\sigma(x)\neq \sigma(y)$ for all adjacent
vertices $x$ and $y$.

\smallskip\par\noindent\textbf{\emph{Rooted Trees}} naturally describe the
phylogenetic relationships both among genes and among species. A rooted
tree $T$ with vertex set $V=V(T)$ and edge set $E=E(T)$ contains a unique
vertex $0_T$, called the \emph{root}, that designates the earliest state
under consideration. Every path in $T$ that originates in $0_T$ thus implies
a temporal order and determines the ancestor-descendant relationship as
follows: If $y$ lies on the unique path from the root to $x \neq y$ then
$y$ is an \emph{ancestor} of $x$, and $x$ is a \emph{descendant} of $y$. In
this case we write $x \prec_{T} y$.  For the edges of $T$ we write $e=xy$
and use the convention that $y\prec_T x$. For an edge $e=xy$, we say that
$x$ is the \emph{parent} of $y$ and $y$ is a \emph{child} of $x$. It will
be useful to extend the ancestor relation to $V(T)\cup E(T)$. For a vertex
$x\in V(T)$ and an edge $e=uv\in E(T)$ we set $x \prec_T e$ if and only if
$x\preceq_T v$; and $e \prec_T x$ if and only if $u\preceq_T x$.  In
addition, for edges $e=uv$ and $f=ab$ in $T$ we put $e\preceq_T f$ if and
only if $v \preceq_T b$.  As usual, $\xi \preceq_{T} \zeta$ is equivalent
to $\xi \prec_{T} \zeta$ or $=\zeta$ for all $\xi,\zeta\in V(T)\cup
E(T)$. If neither $\xi \preceq_{T} \zeta$ nor $\zeta\prec_{T} \xi$ holds,
we say that $\xi$ and $\zeta$ are $\preceq_{T}$-incomparable.  Note that
the root $0_T$ of $T$ is the unique $\prec_T$-maximal vertex.

The set $L(T)$ of all $\prec_T$ minimal vertices form the \emph{leaves} of
$T$.  The \emph{subtree $T(x)$ rooted at $x\in V(T)$} is defined as the
subgraph induced by the vertex set $\{y\in V(T)\mid y\preceq_T x\}$. For
any subset $A\subseteq V(T)$, the \emph{least common ancestor} $\lca(A)$ is
the $\preceq_T$-minimal vertex $w$ that is an ancestor of all $y\in A$.  In
particular, we have $\lca(\{x\})=x$ for all $x\in V(T)$. For simplicity, we
write $\lca(x,y)$ instead of $\lca(\{x,y\})$.  In addition, we may use the
subscript ``$_T$'' to indicate that $\lca_T$ is take w.r.t.\ the tree $T$.
Moreover, we set $\rho_T\coloneqq\lca(L(T))$ and note that $\rho_T=0_T$ if
and only if the root $0_T$ has at least two children.  The set of
\emph{inner vertices} $V^0(T)\coloneqq (V(T)\setminus (L(T)\cup
\{O_T\}))\cup \{\rho_T\}$ of $T$ comprise $V(T)\setminus L(T)$ and excludes
the root if $0_T\ne\rho_T$. A rooted tree is \emph{phylogenetic} if all its
inner vertices have at least two children, and \emph{binary} if all its
inner vertices have exactly two children. We will assume throughout that
$T$ is phylogenetic but not necessarily binary.

\begin{figure}[t]
  \begin{center}
    \begin{tabular}{lcr}
      \begin{minipage}{0.67\textwidth}
        \includegraphics[width=\textwidth]{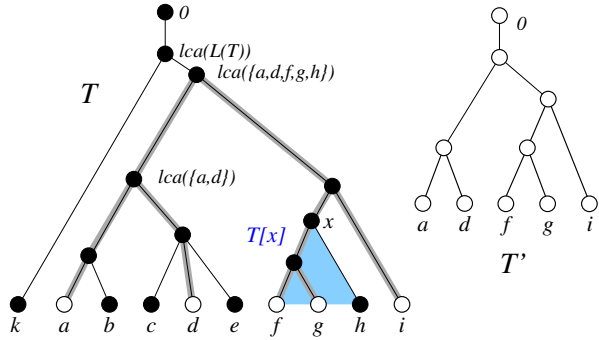}
        \end{minipage} & & 
        \begin{minipage}{0.3\textwidth}
          \caption{A planted phylogenetic tree has root $0$ above the last
            common ancestor $\lca(L(T))$ of all leaves. The tree $T'$ on
            the leaf set $\{a,d,f,g,i\}$ is displayed by $T$. The gray
            outline shows how $T'$ is embedded in $T$. The subtree $T(x)$
            rooted at the inner node $x$ is shown by a blue shading.
            \textit{\small Figure taken from first edition of the this book
            \cite{Setubal:18a}.}
          }
          \label{fig:nomen}
        \end{minipage} 
      \end{tabular}
  \end{center}
\end{figure}

A \emph{refinement} of a phylogenetic tree $T$ is a phylogenetic tree $T'$
on the same leaf set such that $T$ can be obtained from $T'$ by contracting
edges. The \emph{restriction} $T|_{L'}$ of $T$ to $L'$ is the phylogenetic
tree with leaf set $L'$ obtained from the tree $T$ by deleting all vertices
in $L\setminus L'$ and their incident edges and by additionally suppressing
all inner vertices of degree two except the root.  A phylogenetic tree $T'$
on some subset $L'\subseteq L$ is said to be \emph{displayed} by $T$ (or
equivalently that $T$ \emph{displays} $T'$) if $T'$ coincides with
$T|_{L'}$, see Fig.~\ref{fig:nomen}.

\smallskip\par\noindent\textbf{\emph{Rooted triples}} are binary rooted
tree on three vertices. The rooted triple with leaf set and $\lca(x,y)\prec
\lca(x,z)=\lca(y,z)$ will be denoted by $xy|z$.  Triples $xy\vert z$ often
can be derived directly from sequence data and reflect the fact that the
taxa $x$ and $y$ are evolutionary closer related than compared to $z$.
Hence, it is of interest to determine as whether a given set $\R$ of
triples is \emph{consistent}, i.e., there is a tree $T$ that displays all
of the triples in $\R$. Aho et al.\ \cite{Aho:81} devised a polynomial-time
algorithm, called \texttt{BUILD}, that either constructs a uniquely defined
rooted tree $\Aho(\mathscr{R})$ that displays $\R$ or recognizes that $\R$
is inconsistent.  In some situations we also may have information about
triples that are \emph{not} displayed by the tree of interest.  A pair
$(\R,\F)$ of two triple sets is \emph{consistent} if there is a tree $T$
that displays all of the triples in $\R$ but none of the triples in
$\F$. The polynomial-time algorithm \texttt{MTT} (``mixed triplets problem
restricted to trees'') \cite{He:06} tests consistency of pairs $(\R,\F)$ of
triple sets in polynomial time.

\smallskip\par\noindent\textbf{\emph{Planted trees.}}  Evolutionary events
of interest may also pre-date the last common ancestor of all species or
genes. The latter can be accommodated by considering \emph{planted trees},
i.e., trees $T$ that satisfy $0_T\ne\rho_T$. In this case $0_T$ is also
called the planted root. In a planted phylogenetic tree, $0_T$ and
  $\rho_T$ are connected by the edge $0_T\rho_T\in E(T)$.  The planted
root can be thought of a representing the ``outgroup(s)'', while $L(T)$
represents the ``in-group'', i.e., the species or genes under
consideration. A planted tree $T$ always displays the rooted tree
$T(\rho_T)$ obtained by contracting the edge $0_T\rho_T$.

\smallskip\par\noindent\textbf{\emph{Dated Trees.}} 
With each (planted) tree $T$ we can associate a
\emph{time map} $\tau_T:V(T)\to\mathbb{R}$ such that $x\prec_T y$ implies
$\tau_T(x)<\tau_T(y)$. It is an easy task to verify that, for every tree,
such a time map exists. Even more, for every tree one can construct, in
linear-time, a time map such that leaves of the tree are assigned to
particular time-points (e.g.\ all leaves are mapped to the time-point $0$
specifying extant taxa nowadays) \cite{SHL+23}.  It is usually difficult
and often impossible to obtain reliable, accurate ``time stamps''
$\tau_T(x)$ for evolutionary relevant events
\cite{Rutschmann:06,Sauquet:13}. Such detailed information is not needed
for our purposes. The theoretical results will depend only on the existence
of relative timing, i.e., the knowledge as whether an event pre-dates,
post-dates, or is concurrent with another one; an information that is often
much easier to extract from data \cite{Ford:09,Szollosi:22}.

\medskip
\emph{Throughout this chapter we assume that all trees are planted and
phylogenetic, unless explicitly stated otherwise.}

\subsection{Reconciliation}

Genes evolve within genomes, i.e., species. In order to understand the
relationship between genes and species, we need to describe how the gene
tree $T$ ``fits together'' with the species tree $S$. This idea is
formalized by the notion of a \emph{reconciliation map} that locates the
vertices of the gene tree, i.e., the evolutionary events in the history of
the genes, in the species tree.  We model evolutionary events as precise
point in time. This is, of course, an approximation. In reality, even
punctual events such as gene duplications require time to spread through a
population and become fixed. Similarly, speciation is also a
population-based process that takes time \cite{Marques:19} and may also
involve additional effects such as incomplete lineage sorting
\cite{Zheng:14,Chan:17}. Nevertheless, it is justified to view events as
points-in-time in the context of macro-evolutions, where population effects
are neglected altogether.

In order to study reconciliation from a formal point of view, we start
  out with only minimal requirements that are straightforward to argue:
\begin{definition}
  Let $T$ and $S$ be  trees. A \emph{reconciliation} of $T$ and $S$
  is a triple $(T,S,\mu)$ where $\mu: V(T)\to V(S)\cup E(S)$ is a map that
  satisfies the following conditions:
  \begin{itemize}
  \item[\AX{R0}] $\mu(x)=0_S$ if and only if $x=0_T$
  \item[\AX{R1}] $\mu(x)\in L(S)$ if and only if $x\in L(T)$
  \item[\AX{R2}] If $y\prec_T x$ and $\mu(x),\mu(y)\in V(S)$, then
			     $\mu(x)\ne\mu(y)$
  \item[\AX{R3}] If $y\prec_T x$, then $\mu(x)\not\prec_S \mu(y)$
   \end{itemize}
\end{definition}
While Condition \AX{R0} is used to identify the planted roots, Condition
\AX{R1} ensures that leaves of $T$, i.e., extant genes, are found in leaves
of $S$, i.e., in extant species. Condition \AX{R2} ensures that temporally
distinct speciation events cannot be mapped to the same vertex in $S$,
i.e., the same speciation event. Condition \AX{R3}, finally ensures a weak
form of temporal consistency by forbidding that a descendant of $x$ in $T$
can be mapped to an ancestor of $\mu(x)$ in $S$.  Thus, if $y\prec_T x$,
then either $\mu(y)\preceq_S \mu(x)$ or $\mu(x)$ and $\mu(y)$ are
$\preceq_S$-incomparable. For completeness we note that many authors prefer
reconciliation maps of the form $\gamma: V(T) \to V(S)$, i.e., mappings
from vertices to vertices \cite{Tofigh:11,Bansal:12,Stolzer:12}. This can
be easily translated, however: If $\mu(x)=uv$, then $\gamma(x)=v$, i.e.,
everything is mapped to the lower end of the edge in $S$. The other
direction is a bit more intricate. In essence, however, it suffices to set
$\mu(x)=uv$ if $\gamma(x)=v$ and there is a $x'\prec_T x$ with
$\gamma(x')=\gamma(x)=v$.

An important distinction in evolution is the difference between vertical and
horizontal inheritance. Vertical inheritance implies that the the
descendants of a gene are harbored by descendant species. In contrast,
horizontal inheritance, i.e., horizontal transfer consists in the
relocation of a gene to a different lineage. Formally, this situation can
be captured by distinguishing vertical and Horizontal Gene Transfer (HGT)
edges in the gene tree $T$:
\begin{definition}
  \label{def:HGT}
  Let $(T,S,\mu)$ be a reconciliation. An edge $xy\in E(T)$ is an
  \emph{HGT-edge} if $\mu(x)$ and $\mu(y)$ are $\preceq_S$-incomparable.  A
  reconciliation is \emph{HGT-free} if $\mu(x)$ and $\mu(y)$ are
  $\preceq_S$-comparable for all $xy\in E(T)$.
\end{definition}
As a direct consequence of \AX{R3}, a reconciliation $(S,T,\mu)$ is
HGT-free if and only if it satisfies
\begin{itemize}
\item[\AX{R4}]  If $y\prec_T x$ then $\mu(y)\preceq_S \mu(x)$.
\end{itemize}
In fact, \AX{R4} states that all edges of $T$ correspond to vertical
inheritance.

Usually, it is known \emph{a priori} which gene is found in which species.
That is, a map $\sigma: L(T)\to L(S)$ is given that assigns to each extant
gene the extant species in which it occurs.
\begin{definition}
  Let $T$ and $S$ be two  trees and $\sigma: L(T)\to L(S)$. Then
  $(T,S,\mu,\sigma)$ is a $\sigma$-reconciliation if $(T,S,\mu)$ is a
  reconciliation and $\mu_{|L(T)}$ satisfies $\mu_{|L(T)}=\sigma$.
\end{definition}

An early important observation in this field is that every gene tree can be
reconciled with any  species tree even without the need to
consider horizontal gene transfer \cite{Guigo:96,Page:97}:
\begin{theorem}
  \label{thm:exist}
  For any two trees $T$and $S$  and any map $\sigma:L(T)\to L(S)$ there is a HGT-free
  $\sigma$-reconciliation $(T,S,\mu,\sigma)$.
\end{theorem}
To see this, it suffices to consider the reconciliation map $\mu_0$ defined
by setting
\begin{equation}
\begin{split}
&\mu_0(0_T)\coloneqq 0_S; \\ 
&\mu_0(x)\coloneqq \sigma(x) \text{ for all } x\in L(T); \text{ and } \\
&\mu_0(x)\coloneqq 0_S\rho_S \text{ for all inner vertices } x\in V^0(T).
\end{split}
\label{eq:mu0}
\end{equation}

\begin{figure}[t]
  \begin{center}
    \begin{tabular}{lcr}
      \begin{minipage}{0.47\textwidth}
        \includegraphics[width=\textwidth]{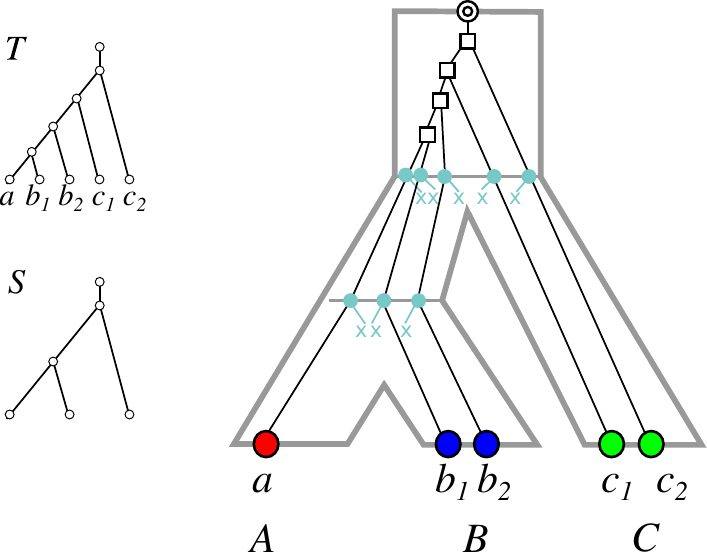}
      \end{minipage} & $\quad$	 & 
      \begin{minipage}{0.5\textwidth}
        \caption{Illustration of the $\sigma$-reconciliation
          $\scen=(T,S,\mu_0,\sigma)$ with $\mu_0$ defined by
          Eq.~\eqref{eq:mu0}.  Genes are colored w.r.t.\ the species in
          which they reside, i.e., $\sigma(a)=A$ (red),
          $\sigma(b_1)=\sigma(b_2)=B$ (blue), and
          $\sigma(c_1)=\sigma(c_2)=C$ (green).  Note that the gene tree $T$
          and the \emph{observable part} of the gene tree in
          Fig.\ \ref{fig:trueHist} which is obtained by removing the
          $\times$-edge and suppression of its two incident vertices,
          coincide.  Fig.\ \ref{fig:trueHist} and the present figure thus
          show different reconciliations for the same trees $T$ and $S$. In
          particular, the reconciliations feature very different numbers of
          gene losses: a single one in Fig.\ \ref{fig:trueHist} and $8$ for
          $\mu_0$.  }
        \label{fig:muh-zero}
      \end{minipage} 
    \end{tabular}
  \end{center}
\end{figure}

Every reconciliation map $\mu$ can be used to imply \emph{gene loss}
events. To see this, consider an edge $xy\in V(T)$ such that
$\mu(y)\prec_S\mu(x)$ for a given reconciliation map $\mu$ and for which
there is a vertex $u\in V(S)$ with $\mu(y)\preceq_S u\preceq_S\mu(x)$, and
thus $u\notin L(S)$. During the speciation $u\in V^0(S)$, a descendant of
the each gene present in the genome is transmitted to every descendant
lineage, i.e., to every edge $uw\in E(S)$. If no descendant of the gene is
found in a species descending from $w$, then the gene must have died out in
the entire species subtree $S(w)$. This is explained most parsimoniously by
a single loss event occurring already along the edge $uw\in
E(S)$. Fig.~\ref{fig:muh-zero} shows that for the case of the
reconciliation $\mu_0$ this reasoning implies an unrealistically large
number of gene loss events: for example, the five paralogs present at the
earliest speciation give rise to a total of 10 genes, of which half are
subsequently lost again. This reconciliation $\mu_0$ will in general not be
a plausible biological explanation because it implies unreasonably large
number of gene losses. This observation led to the development of a number
of alternative scoring functions $F(\mu)\coloneqq F(T,S,\mu,\sigma)$ for
reconciliation maps with given trees $T$ and $S$ and a given leaf
assignment $\sigma$ that count the number of duplication and/or gene
losses.  We refer to \cite{Zhang:97} and the references therein for
detailed discussion.

We briefly consider this optimization problem for the HGT-free case here.
Assume that we are given a leaf-assignment $\sigma$. The task is then to
find a $\sigma$-reconciliation that optimizes a given scoring function. By
Thm.~\ref{thm:exist} we know that a solution always exists. Since any
vertex $x\in V(T)$ has to be mapped at or above the last common ancestor of
all the species in which the descendants of $x$ are found, every solution
satisfies
\begin{equation} \label{eq:x-above-species}
\lca_S(\sigma(L(T(x)))) \preceq_S \mu(x)
\end{equation}
The so-called \emph{LCA reconciliation} $\mu^*$ attains equality whenever
possible. More precisely, we define $\mu^*$ as follows. For each $x\in
V(T)$, let $w(x)\coloneqq \lca_S(\sigma(L(T(x))))$ and $u$ be the 
parent of $w(x)$ in $S$ which always exists in planted trees.
\begin{equation*}
\mu^*(x)\coloneqq \begin{cases}
  w(x)  & \text{whenever there is no $y$ with $y\prec_T x$ and $w(x)=w(y)$} \\
  u w(x) & \text{otherwise.}
\end{cases}
\end{equation*}

The LCA-reconciliation $\mu^*$ is the most parsimonious reconciliation
w.r.t.\ to several cost measures including the number of duplications and
the number duplications and gene losses
\cite{Chen:00,Zhang:97,Zmasek:01}. Corresponding maximum likelihood and
Bayesian methods is described in \cite{Gorecki:11} and \cite{Arvestad:03},
respectively.  Most parsimonious reconciliations can still be computed in
polynomial time if HGT events are included and penalized
\cite{Bansal:12,Tofigh:11}. However, as we shall see in the following
section, these reconciliations are not always biologically feasible,
see also \cite{Menet:22}.

\subsection{Relaxed Scenarios} 

It was noted in \cite{Bansal:12,Tofigh:11} that the definition of
reconciliations in the previous section is not sufficient to ensure that
$(T,S,\mu)$ can be interpreted as series of events along a linear time
axis. To this end we consider evolutionary scenarios as reconciliations of
dated trees:
\begin{definition}
  \label{def:rs}
  A \emph{(relaxed) scenario} $\scen=(T,S,\mu,\tau_T,\tau_S)$ consists of a
  dated gene tree $(T,\tau_T)$, a dated species tree $(S,\tau_S)$, and a
  \emph{reconciliation map} $\mu \colon V(T)\to V(S)\cup E(S)$ such that
  \begin{itemize}
  \item[\AX{S0}] $\mu(x)=0_S$ if and only if $x=0_T$,
  \item[\AX{S1}] $\mu(x)\in L(S)$ if and only if $x\in L(T)$,  
    \item[\AX{S2}] $\tau_S(\mu(x))=\tau_T(x)$ for all $x\in V(T)$ with
      $\mu(x)\in V(S)$, and 
    \item[\AX{S3}] $\tau_S(v)<\tau_T(x)<\tau_S(u)$ for all $x\in V(T)$ with
      $\mu(x)=uv\in E(S)$.
  \end{itemize}
\end{definition}
Axioms \AX{S0} and \AX{S1} coincide with \AX{R0} and \AX{R1}, respectively,
in the definition of a reconciliation.  The remaining two axioms, \AX{S2}
and \AX{S3}, specify \emph{time consistency}. As for reconciliations in the
previous section, we do not assume that a leaf-map $\sigma:L(T)\to L(S)$ is
given, although this typically will be the case in practical applications.
\begin{definition}
  A tuple $(T,S,\mu,\tau_T,\tau_S,\sigma)$ is is a \emph{relaxed
  $\sigma$-scenario} if $(T,S,\mu,\tau_T,\tau_S)$ is a relaxed scenario
  and $\sigma:L(T)\to L(S)$ satisfies $\sigma=\mu_{|L(T)}$.
\end{definition}

Relaxed scenarios can be seen as reconciliations for which a time axis is
given explicitly. It is not difficult to show that for a given a relaxed
scenario $(T,S,\mu,\tau_T,\tau_S)$, the triple $(T,S,\mu)$ is indeed a
reconciliation (cf.\ Lemma 2 and 3 in \cite{SHL+23}).  The converse,
however, is not true in general, i.e., it is not always possible to find
time stamps that turn a reconciliation into a relaxed scenario
cf.\ e.g.\ Fig.\ 2 in \cite{Nojgaard:18a}.
This motivates the following
\begin{definition}
  A reconciliation $(T,S,\mu)$ is \emph{time consistent} if there are
  dating maps $\tau_T$ and $\tau_S$ such that $(T,S,\mu,\tau_T,\tau_S)$ is
  a relaxed scenario.
\end{definition}

\begin{figure}[t]
  \centering
  \includegraphics[width=\textwidth]{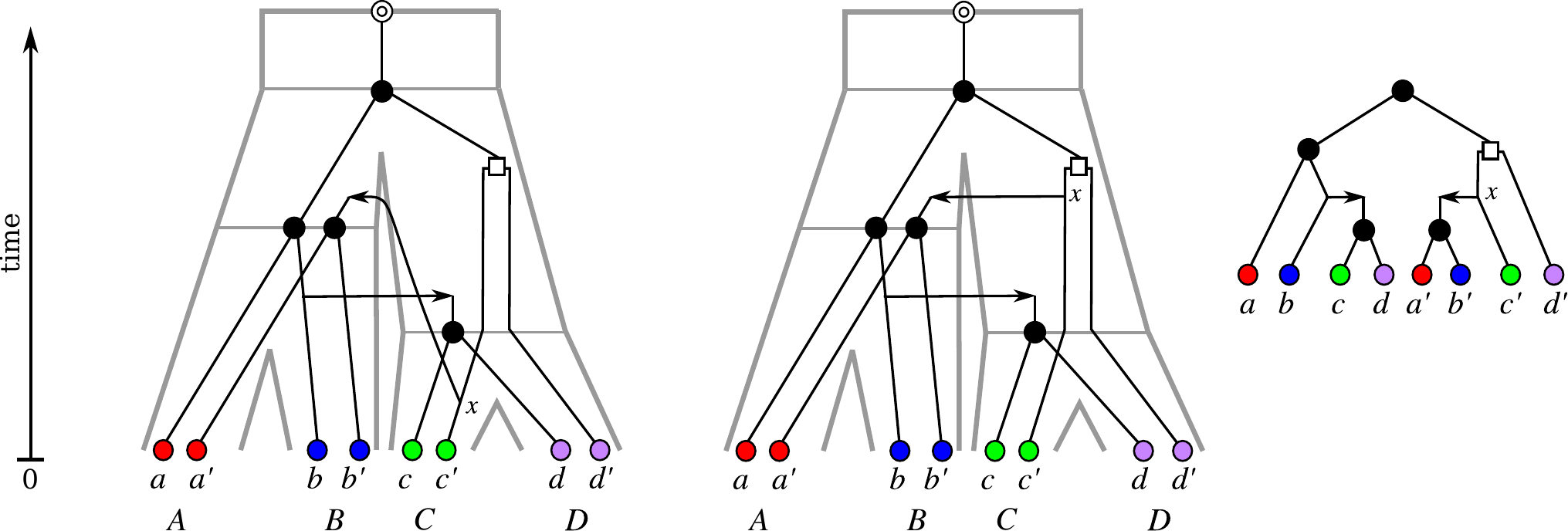}
  \caption{An event-annotated gene tree (left) where we replaced the
    event-labels $\LEAF$ of the leaves $v$ by vertices with color
    $\sigma(v)$ to indicate in which species $v$ resides.  In particular,
    we have $\sigma(x)=\sigma(x')=X$ for $(x,x',X) \in \{ (a,a',A) ,
    (b,b',B) , (c,c',C) , (d,d',D) \}$.  Shown are two reconciliations into
    the same species tree $S$ (left and middle).  The reconciliation shown
    in the middle is time-consistent and, in particular, results in a
    relaxed scenario.  Let us denote with $\mathsf c$ and $\mathsf d$ the
    species that contain $c,c'$ and $d,d'$, respectively and put
    $u\coloneqq \lca_S(\mathsf c, \mathsf d)$ The two reconciliations
    differ only in the choice of placing the origin $x\in V(T)$ of the
    HGT-edge either on the edge $u\mathsf c$ or on the edge $\rho_Su$.  In
    the first case, however, we have $y\prec_T x$ but $t_T(x)<t_T(y)$.
    Hence, the reconciliation shown right does not give rise to a valid
    time map of $T$ and is, therefore, not time-consistent.  }
  \label{fig:timeCons}
\end{figure}

A variety of different auxiliary graphs on $V(S)$
\cite{Tofigh:11,Stolzer:12} or $V(S)\cup V(T)$
\cite{Nojgaard:18a,Lafond:20} were introduced to summarize the temporal
constraints. In each case, it was shown that a reconciliation $(T,S,\mu)$
is time consistent if and only if the corresponding auxiliary graph is
acyclic. It can be shown, furthermore, that these auxiliary graphs are
acyclic in the special case of HGT-free reconciliations, i.e., every
HGT-free reconciliation is automatically time-consistent and can be
extended to a relaxed scenario.

In the general case, the additional requirement of time-consistency also
makes the computation of maximum parsimony reconciliations difficult. As
shown in \cite{Tofigh:11}, the problem then becomes NP hard. 

\subsection{Event-Labels}

From a biological point of view, the inner vertices $V^0(T)$ of the gene
tree $T$ model evolutionary events. In the case of HGT-free scenarios,
i.e., reconciliations $(T,S,\mu)$ that satisfy \AX{R4}, there are only two
fundamentally distinct events: gene duplications and speciation events.
Since speciation events are also modeled by the inner vertices $V^0(S)$, it
is natural to distinguish duplications and speciation by their image under
the reconciliation map $\mu$. This naturally leads to the following
definition \cite{Geiss:20b}:

\begin{definition} 
  \label{def:events}
  Let $(T,S,\mu)$ be a HGT-free reconciliation. Then the 
  \emph{event labeling on $T$} is the map
  $\ell_{\mu}:V(T)\to \{\ROOT,\LEAF,\SPEC,\DUPL\}$ given by:
  \begin{equation*}
  \ell_\mu(u) = \begin{cases}
  \ROOT & \, \text{if } u=0_T \text{, i.e., } \mu(u)=0_S\\
  \LEAF & \, \text{if } u\in L(T) \text{, i.e., } \mu(u)\in L(S)\\
  \SPEC & \, \text{if } \mu(u)\in V^0(S)  \text{ (speciation)}\\
  \DUPL & \, \text{else, i.e., } \mu(u)\in E(S) \text{ (duplication)}\\
  \end{cases}
  \end{equation*}
  \label{def:event-rbmg}
\end{definition}
The first two cases, $\ROOT$ and $\LEAF$, distinguish the planted root and
the leaves of $T$. The remaining two cases identify \emph{speciation}
events with the $u\in V(T)$ that are mapped to inner vertices in the
species tree. In contrast, if $\mu(u)\in E(T)$, it does not correspond to a
speciation and thus corresponds to a gene \emph{duplication}, denoted by
$\DUPL$.  The conditions on the reconciliation $(T,S,\mu)$ considered so
far, however, are not sufficient to ensure that $x\in V^0(T)$ with
$\ell_{\mu}(x)=\SPEC$ represents a biologically meaningful speciation
event. Additional constraint on the evolutionary scenario need to be
introduced. 

This is possibly best explained by assuming to have a binary gene tree $T$.
In this case, if $\ell_{\mu}(x)=\SPEC$, the two children $v'$ and $v''$ of
$x$ should be mapped via $\mu$ into two distinct lineages. Otherwise, if
$v'$ and $v''$ are mapped into the same lineage there is no clear
historical trace that justifies $x$ to be a speciation vertex.  In other
words, $\ell_{\mu}(x)=\SPEC$ is biologically plausible, if for the two
children $v'$ and $v''$ of $x$ the images $\mu(v')$ and $\mu(v'')$ are
$\preceq_S$-incomparable.  Moreover, similar to
Eq.\ \ref{eq:x-above-species} and to accommodate most parsimonious
reconciliation (LCA-reconciliation) one may assume that
$\mu(x)=\lca_S(\mu(v'),\mu(v''))$.  The latter discussion naturally
translates to the case of general not necessarily binary gene trees as
follows: 
\begin{itemize}
\item[\AX{R5}] Suppose $\mu(x) \in V(S)^0$ for some $x\in V$. Then
  \begin{itemize}
  \item[(i)] $\mu(x)=\lca_S(\mu(v'),\mu(v''))$ for at least two distinct
    children $v',v''$ of $x$ in $T$.
  \item[(ii)] $\mu(v')$ and $\mu(v'')$ are incomparable in $S$ for any two
    distinct children $v'$ and $v''$ of $x$ in $T$.
  \end{itemize}
\end{itemize}
As shown in \cite{Geiss:20b,Nojgaard:18a}, the axioms \AX{R0}-\AX{R5} are
equivalent to axioms that are commonly used in the literature
\cite{Gorecki:06,Vernot:08,Doyon:11,Rusin:14,Hellmuth:17}.  In particular,
for any HGT-free $\sigma$-reconciliation $(T,S,\mu,\sigma)$, there is a
LCA-reconciliation for $T$ and $S$ \cite{Hellmuth:17}.  The latter
observation can easily be extended for $\sigma$-reconciliations involving
HGTs and by considering LCA-maps restricted to the HGT-free subtrees of $T$
\cite{Hellmuth:17}.  Thus, the axiom set used here naturally corresponds to
LCA-mappings and hence, to most parsimonious reconciliations. In addition,
\cite[Lem.2]{Geiss:20b} shows that $L(T(v'))\cap L(T(v''))\ne\emptyset$ for
a pair of distinct children $v'$ and $v''$ of $x$ implies
$\ell_{\mu}(x)=\DUPL$ for all HGT-free $\sigma$-reconciliations that
satisfy \AX{R5}.  This is turn motivates to consider the \emph{extremal
event labeling} $\hat \ell$, which assume that these are the only
duplications and thus assigns $\hat \ell(x)=\SPEC$ whenever
$\mu(L(T(v')))\cap\mu(L(T(v''))=\emptyset$ for all pairs of children of
$x$. It is important to note that the extremal labeling $\hat \ell$ is
defined solely on the information of $T$ and does not depend on the
existence of a species tree $S$ or a reconciliation map $\mu$.There
  is no \textit{a priori} guarantee, therefore, that the extremal event
  labeling can be realized by an actual biological scenario.

\section{Best Match Graphs and Orthology}

\subsection{Definition and Characterization}

Many of the combinatorial methods for determining orthology start from
reciprocal best (\texttt{blast}) hits. Here, we consider \emph{best
matches} as a basic evolutionary concept that is approximated on sequence
data by ``best hits''. We therefore consider best matches relative to an
underlying phylogenetic tree, albeit this tree is usually unknown. Our
starting point is therefore a gene tree $T$ together with a map $\sigma:
V(T) \to Y$, where $Y$ is a set of species. It will be convenient to treat
$\sigma$ as a coloring of the leaves of tree $T$ by the species in which
the extant genes reside.

\begin{definition}\label{def:(R)BMG}
  Let $(T,\sigma)$ be a leaf-colored tree. A leaf $y\in L(T)$ is a
  \emph{best match} of the leaf $x\in L(T)$ if $\sigma(x)\neq\sigma(y)$ and
  $\lca(x,y)\preceq_T \lca(x,y')$ holds for all leaves $y'$ from species
  $\sigma(y')=\sigma(y)$.  The leaves $x,y\in L(T)$ are \emph{reciprocal
    best matches} if $y$ is a best match for $x$ and $x$ is a best match
  for $y$.
\end{definition}
Note that a gene $x$ may have two or more (reciprocal) best matches of the
same color $r\neq \sigma(x)$. Some orthology detection tools, such as
\texttt{ProteinOrtho} \cite{Lechner:11a}, explicitly attempt to extract all
reciprocal best matches from the sequence data.

A directed, vertex-labeled graph $(G,\sigma)$ is a \emph{best match} graph
(BMG) if there is a leaf-labeled tree $(T,\sigma)$ such that $(x,y)$ is a
directed edge in $(G,\sigma)$ if and only if $y$ is best match of $x$ in
$(T,\sigma)$.  Given $(T,\sigma)$ we write $\BMG(T,\sigma)$ for its best
match graph.

A key observation towards characterizing best match graphs is that some
subsets of vertices on two colors (species) yield constraints on triples
displayed by any leaf-colored tree that might explain a BMG.
\begin{definition}\label{def:informative_triples}
  Let $a,b,b'$ be pairwise distinct vertices in a colored digraph
  $(G,\sigma)$ such that $\sigma(a)\neq\sigma(b)=\sigma(b')$ and $(a,b)\in
  E(G)$. Then the triple $ab|b'$ is \emph{informative} if $(a,b')\notin
  E(G)$ and \emph{forbidden} if $(a,b')\in E(G)$.
\end{definition}
The sets $\R(G,\sigma)$ and $\F(G,\sigma)$ denotes the set of all
informative and forbidden triples, respectively.  It is shown in
\cite{Geiss:20b} that, if $(G,\sigma)=\BMG(T,\sigma)$, then $T$ displays
all triples in $\mathscr{R}(G,\sigma)$ and none of the triples in
$\mathscr{F}(G,\sigma)$. The sets $\mathscr{R}(G,\sigma)$ and
$\mathscr{F}(G,\sigma)$ also give rise to convenient characterizations of
BMGs.

\begin{thm}
  \label{thm:BMG}
  A properly colored digraph $(G,\sigma)$ is a BMG if and only if one of
  the following conditions is satisfied.
  \begin{itemize}
  \item[(i)] $\mathscr{R}(G,\sigma)$ is consistent and
    $\BMG(\Aho(\mathscr{R}(G,\sigma)),\sigma)=(G,\sigma)$
    \cite{Geiss:20b,Schaller:21d}
  \item[(ii)] $(\mathscr{R}(G,\sigma),\mathscr{F}(G,\sigma))$ is consistent
    and $(G,\sigma)$ is color-sinkfree, i.e., for every $x\in V(G)$ and
    every color $s\ne\sigma(x)$ there is a vertex $y$ such that $(x,y)\in
    E(G)$ and $y\in N(x)$ with $\sigma(y)=s$.  \cite{Schaller:21b}
  \end{itemize}
  In particular, BMGs be be recognized in polynomial-time.
\end{thm}

Intriguingly, every BMG $(G,\sigma)$ is associated with a \emph{unique}
least-resolved tree $(T^*,\sigma)$. That is, (i)
$\BMG(T^*,\sigma)=(G,\sigma)$ and (ii) if $\BMG(T,\sigma)=(G,\sigma)$ then
$T$ displays $T^*$. The least resolved tree $T^*$ therefore captures the
reliable phylogenetic information about the gene tree $T$ that is provided
by the best matches. The least-resolved tree for a BMG $(G,\sigma)$ is
precisely the tree $\Aho(\mathscr{R}(G,\sigma))$ and can be constructed in
polynomial time.

While least-resolved trees serve as a scaffold to cover phylogenetic
information without making more assumptions on the evolutionary history
than actually provided by the data, one is in many cases additionally
interested to find binary trees which can be considered as the most
``highly'' resolved histories.  Binary trees are of particular interest
because true multifurcations are most likely rare, i.e., most polytomies
are a consequence of insufficient resolution of the available data
\cite{Maddison:89,DeSalle:94,Walsh:99}.  However, not every BMG can be
explained by a binary tree (cf.\ \cite[Fig.6A]{Schaller:21a}). Binary
explainable BMGs are characterized as those BMGs that do not contain a
certain colored graph on four vertices, termed \emph{hourglass}, as induced
subgraph \cite{Schaller:21a}. Note that binary trees explaining a BMG are
not necessarily unique, however, they all display
$\Aho(\mathscr{R}(G,\sigma))$ and can be constructed in polynomial time
\cite{Schaller:21e}.

Binary explainable BMGs also have a convenient characterization in terms of
triple sets. Consider the following extension of the set of informative
triples:
\begin{equation}
  \mathscr{R}^B(G,\sigma) \coloneqq  \mathscr{R}(G,\sigma) \cup
	\left\{ bb'|a : ab|b'\in\mathscr{F}(G,\sigma) \text{ and }
  \sigma(b)=\sigma(b')\right\}
\end{equation}
As shown in \cite{Schaller:21e}, replacing $\mathscr{R}(G,\sigma)$ by
$\mathscr{R}^B(G,\sigma)$ in Thm.~\ref{thm:BMG} yields a characterization
of binary explainable BMGs. Moreover, if $(G,\sigma)$ is a binary
explainable BMG, then $T^B \coloneqq \Aho(\mathscr{R}^B(G,\sigma))$ has the
property that a binary tree $(T,\sigma)$ satisfies
$\BMG(T,\sigma)=\BMG(T^B,\sigma)=(G,\sigma)$ if and only if $T$ is a
refinement of $T^B$. We summarize these results in the following
\begin{thm} 
  The following statements are equivalent for every BMG $(G, \sigma)$:
  \begin{itemize}%[noitemsep]
  \item[(i)] $(G,\sigma)$ is binary explainable. 
  \item[(ii)] $(G,\sigma)$ is hourglass-free \cite{Schaller:21a}. 
  \item[(iii)] $\mathscr{R}^B(G,\sigma)$ is consistent \cite{Schaller:21e}.  
  \end{itemize}	
  In this case, a binary tree that explains $(G,\sigma)$ can be constructed
  in polynomial time. In particular, the BMG $(G, \sigma)$ is explained by
  every refinement of the tree $(\Aho(\mathscr{R}^B(G,\sigma)), \sigma)$
  \cite{Schaller:21e}.
\end{thm}

In practice, $(G,\sigma)$ is obtained empirically by comparing similarities
of gene sequences. Most likely, $(G,\sigma)$ thus will not be a BMG but
differ from the true best match graph by both false positive and false
negative edges. The arc modification problems for BMGs aim at correcting
such errors. Like most graph editing problems, it is NP-complete
\cite{Schaller:21b,Schaller:21e}.  However, efficient heuristics can be
devised that solve the problem with acceptable accuracy \cite{Schaller:21g}

\subsection{Orthology in the Absence of HGT}

Walter Fitch \cite{Fitch:70} defined orthology as homology deriving from a
speciation event. While later discussions qualified this simple concept in
the presence of HGT (see below), the notion is clear in a HGT-free setting.
\emph{We assume throughout this section that scenarios and reconciliations
are HGT-free.}
\begin{definition}\label{def:ortho-para}
  Let $(T,t)$ be an event-labeled gene tree and let $x,y\in L(T)$ be two
  distinct genes. Then $x$ and $y$ 
  are \emph{orthologs} if $t(\lca_T(x,y))=\SPEC$; they are \emph{paralogs}
  if $t(\lca_T(x,y))=\DUPL$. 
\end{definition}
We write $\Theta(T,t)$ for the \emph{orthology graph} with vertex set
$L(T)$ and edges $xy\in E(\Theta(T,t)$ whenever $x$ and $y$ are orthologs.
Note that the corresponding ``paralogy graph'' is simply the complement
graph of $\Theta(T,t)$. Moreover, $\Theta(T,t)$ is symmetric but not
necessarily transitive (edges $xy$ and $yz$ do not imply that $xz$ is an
edge).  Orthology graphs feature a simple structure:
\begin{thm}[\cite{Hellmuth:13a}]\label{thm:ortho-cograph}
  A graph $G$ is an orthology graph for some event-labeled tree $(T,t)$,
  i.e. $G = \Theta(T,t)$, if and only if $G$ is a cograph.
\end{thm}
A classical characterization of cographs is the following: $G$ is a cograph
if and only if it does not contain a $P_4$, i.e, a path on four vertices,
as an induced subgraph \cite{Corneil:81}.

Given a reconciliation $(T,S,\mu)$, orthology is therefore implied by the
reconciliation maps, since, by Def.~\ref{def:events} above, $\mu$ specifies
the event labeling $\ell_{\mu}$.  The dependence of orthology on $\mu$ is
crucial. In the extreme case of $\mu_0$, we have $\mu_0(V(T^0))\subseteq
E(S)$, and thus all events are classified as duplications. Conversely, the
LCA reconciliation maps as many $u\in V^0(T)$ to inner vertices of $S$ as
possible, and thus can be expected to result in a large number orthologous
pairs.

The starting point for many algorithmic approaches to orthology detection
is the observation that two orthologs $x$ and $y$ are also reciprocal best
matches. We are therefore in particular interested in obtaining information
on the orthology relation starting from best match data only.

Writing $\BMG_{sym}(T,\sigma)$ for the subgraph of the best match
$\BMG(T,\sigma)$ comprising the reciprocal, i.e., bi-directional edges, and
denoting by $\hat\ell_T$ the extremal event labeling for $T$ introduced in
the previous section, we obtain the following key result:
\begin{thm}[\cite{Geiss:20b}] \label{thm:orthoSubsetRBMG}
  Let $(T,S,\mu,\sigma)$ be a HGT-free $\sigma$-reconciliation satisfying
  \AX{R5}. Then $\Theta(T,\ell_{\mu})\subseteq \Theta(T,\hat\ell_T)\subseteq
  \BMG_{sym}(T,\sigma)\subseteq \BMG(T,\sigma)$.
\end{thm}
Theorem \ref{thm:orthoSubsetRBMG}, in particular, shows that using the
reciprocal best match graph $\BMG_{sym}(T,\sigma)$ as an approximation for
the orthology relation does not yield false negative orthology
assignments. In general, however, $\Theta(T,\ell_{\mu})$ is a proper
subgraph of $\BMG_{sym}(T,\sigma)$, i.e., reciprocal best match graphs may
contain false-positive orthology assignments.  In particular,
$\BMG_{sym}(T,\sigma)$ does not need to be a cograph and thus may contain
induced $P_4$s.

If $L(T(v'))\cap L(T(v''))\ne\emptyset$ for some children $v'$ and $v''$ of
$x$ then $\ell_{\mu}(x)=\DUPL$ for all HGT-free $\sigma$-reconciliations
that satisfy \AX{R5} (cf.\ \cite[Lem.2]{Geiss:20b}).  \cite[Lemma
  10]{Schaller:21a} characterizes false orthology assignments in best
matches graphs and provides an even stronger results.  In particular, it
shows that $\hat\ell_{\mu}(x)=\DUPL$ precisely if $L(T(v'))\cap
L(T(v''))\ne\emptyset$ for some children $v'$ and $v''$ of $x$. Using the
fact that $\Theta(T,\ell_{\mu})\subseteq \Theta(T,\hat\ell_T)$ allows us to 
rephrase this result in following, more convenient form: 
\begin{thm}\label{thm:char-para}
  The following statements are equivalent for every tree $(T,\sigma)$ and
  any two leaves $x$ and $y$ of $T$.
\begin{itemize}
\item[(i)] $x$ and $y$ are paralogs in \emph{any} HGT-free
  $\sigma$-reconciliation $(T,S,\mu,\sigma)$
\item[(ii)] There are two children $v'$ and $v''$ of $\lca_T(x, y)$ such
  that $L(T(v'))\cap L(T(v''))\ne\emptyset$.
\item[(iii)] $xy$ are ``false orthologs'' in case $x$ and $y$ are
  reciprocal best matches.
\end{itemize}
\end{thm}
Note that Thm.\ \ref{thm:char-para} depends on the structure of the
underlying tree $(T,\sigma)$ and trees that explain a given BMG are not
necessarily unique.  Hence, there might be a tree $(T',\sigma)$ such that
$\BMG(T,\sigma)= \BMG(T',\sigma)$ but for which an edge $xy$ is a
false-positive orthology assignment w.r.t.\ $(T,\sigma)$ but not
w.r.t.\ $(T',\sigma)$. Hence, we are interested, in particular, in those
bi-directional edges $xy$ of a BMG $(G,\sigma)$ that are false orthology
assignment for \emph{every} tree $(T,\sigma)$ that satisfies
$\BMG(T,\sigma)=(G,\sigma)$.  Such edges are called \emph{unambiguously}
false orthology assignment in $(G,\sigma)$ and can be identified in chains
of overlapping hourglasses. For the details we refer to Def.14 and Def.16
in \cite{Schaller:21a} for the detailed specification of all ``hug''-pairs
and we obtain
\begin{thm}[{\cite[Thm.11 \& 12]{Schaller:21a}}]\label{thm:ufp}
  An edge $xy$ in a BMG $(G,\sigma)$ is an unambiguously false orthology
  assignment if and only $x$ and $y$ is a ``hug''-pair.  The set of all
  unambiguously false orthology assignment in a BMG can be computed in
  polynomial time.
\end{thm}

The \emph{no-hug graph} $\mathfrak{N}(G,\sigma)$ is the subgraph of
$(G_{sym},\sigma)\subseteq (G,\sigma)$ from which all edges $xy$ that are
hug pairs have been removed.  In particular, $\mathfrak{N}(\BMG(T,\sigma))$
contains the orthology graph for every HGT-free $\sigma$-reconciliation
satisfying \AX{R5} $\mu$ as a subgraph (cf.\ \cite[Cor.\ 5]{Schaller:21a}):
\begin{equation} 
  \Theta(T,\ell_{\mu})\subseteq \Theta(T,\hat \ell_T))\subseteq
  \mathfrak{N}(\BMG(T,\sigma)) \subseteq \BMG_{sym}(T,\sigma)
\end{equation}
As shown in \cite{Schaller:21a}, $\mathfrak{N}(\BMG(T,\sigma))$ is in fact
an orthology graph. However, since only unambiguously false orthology have
been removed, $\mathfrak{N}(\BMG(T,\sigma))$ may still contain false
orthologs. That is, reciprocal best matches $xy$ might be paralogs in the
true scenario $(T,\ell)$ but appear as orthologs in alternative scenarios
that are consistent with the best match data. In fact, it is always
possible to ``move up'' a speciation event and to replace it by a
duplication followed by losses. In this manner it is always possible to
explain best match without orthologs. The no-hug graphs
$\mathfrak{N}(\BMG(T,\sigma))$ thus provides the ``most parsimonious''
explanation in the sense that it predicts an orthology relationship
whenever this is consistent with the best match data instead of an
alternative explanation, which would comprise a duplication event
accompanied by complementary losses. Thus, the reconciliation underlying
$\mathfrak{N}(\BMG(T,\sigma))$ is a least conceptually related to LCA
reconciliations. A closer inspection of this connection, however, is a
topic for future research.

In practice, one could obtain an accurate orthology assignment by
estimating an initial, species labeled graph $(\tilde G,\sigma)$
representing best (\texttt{blast}) hit data and then edit these initial
data to a BMG, i.e., a graph $(G,\sigma)$ that conforms to
Def.\ \ref{def:(R)BMG}.  This step is, in general, NP-hard
\cite{HGS:20,Schaller:21b} and thus, will require an efficient
heuristic. From the the BMG $(G,\sigma)$ we can then remove, in
polynomial-time, all hug-pairs to obtain the no-hug graph
$\mathfrak{N}(G,\sigma)$ that is free of unambiguously false orthologs and
also an orthology graph.
  
Since orthology graphs are cographs, there is also a more direct, albeit
less accurate approach towards estimating the orthology graph.  To this
end, one extracts the reciprocal best hits $(\tilde G_{sym},\sigma)$ from
the initial estimated best (\texttt{blast}) hits $(\tilde G,\sigma)$ and
finds the cograph $H$ that differs by the fewest edge-insertions and
edge-deletions from $\tilde G_{sym}$. This cograph editing problem is also
NP-hard \cite{Liu:12}. However, it remains tractable if $\tilde G_{sym}$ is
not too dissimilar from a cograph. Fast heuristics have become available
for this task in last years \cite{White:18,Hellmuth:20b,Crespelle:21}.

By Theorem \ref{thm:BMG}, BMGs can be recognized in polynomial-time and the
tree $(T_{\Aho},\sigma) \coloneqq (\Aho(\mathscr{R}(G,\sigma)),\sigma)$ is
the unique least resolved tree that explains a BMG $(G,\sigma)$. However,
it is not possible in general to equip $T_{\Aho}$ with an event labeling
$\ell$ such that $\Theta(T_{\Aho},\ell)=\mathfrak{N}(G,\sigma)$. Instead,
$T_{\Aho}$ can be augmented to a uniquely defined tree $\hat\ell_{T^*}$
that is equipped with the extremal event labeling $\hat\ell_{T^*}$.  The
event labeled tree $(T^*,\hat\ell_{T^*})$ satisfied
\begin{equation}
  \Theta(T^*,\hat \ell_{T^*}) =  \mathfrak{N}(\BMG(T,\sigma))
\end{equation}
and can be obtained in  polynomial time (cf.\ \cite[Thm.10]{Schaller:21a}).
The extremal labeling $\hat \ell_{T^*}$ is defined solely on the
information of $T^*$ and does not depend on the existence of a species tree
$S$ or a reconciliation map $\mu$.

In summary, therefore, it is possible to use the information of a BMG
$(G,\sigma)$ to get mathematically sound estimates $G'$ of orthology
relations and a resulting gene tree $(T,\ell)$ such that $G' =
\Theta(T,\ell)$ in polynomial time.

An intriguing observation is that the event labeled gene trees $(T,\ell)$
have implications for the species tree irrespective of the details of the
reconciliation map. To this end, let $\Striple(T,\ell,\sigma)$ denote the
set of all triples $\sigma(x)\sigma(y)|\sigma(z)$ that satisfy (i)
$\sigma(x)$, $\sigma(y)$, and $\sigma(z)$ are pairwise distinct species and
(ii) $\ell(\lca(x,z))=\SPEC$, i.e., the root of $xy|z$ is a speciation
event.
\begin{proposition}[\cite{HernandezRosales:12a,Hellmuth:17}]
  \label{prop:species-triples-ortho}
  For a given event-labeled tree $(T,\ell)$ and species map $\sigma$ the
  following two statements are equivalent.
 \begin{itemize}
 \item[(i)] There is a species tree $S$ such that there is a HGT-free
   $\sigma$-reconciliation $(T,S,\mu,\sigma)$ with $\ell = \ell_\mu$.
 \item[(ii)] The triple set $\Striple(T,\ell,\sigma)$ is consistent. 
 \end{itemize}
\end{proposition}
Note that the set of species triples in
Prop.~\ref{prop:species-triples-ortho} depends on the event-labeling $\ell$
of $T$, but not on the details of the reconciliation map $\mu$ as long as
$\mu$ gives rise to correct event-labeling.  Several types of problems that
are concerned with optimally editing a given undirected graph to an
orthology relation that, at the same time, satisfies additional constraints
(e.g.\ that the resulting event-labeled gene tree can be reconciled with
some (unknown) species tree) have been considered e.g.\ in
\cite{Lafond:16}. While most of the latter problems are NP-complete,
certain types of problems that are related to the correction of homology
relations that provide only partial information about orthologs and
non-orthologs can be solved in polynomial-time
\cite{Lafond:14,Nojgaard:18b}.

Although Prop.~\ref{prop:species-triples-ortho} is rather technical, it has
significant practical importance. An estimate of $(T,\ell)$, e.g.\ obtained
directly from reciprocal best match data by means of cograph editing
provides a collection of species triples.  Pooling these data over a large
number of gene families indeed yields sufficient information to infer fully
resolved species trees \cite{Hellmuth:15a}.

\begin{figure}[t]
  \centering
  \includegraphics[width = 0.9\textwidth]{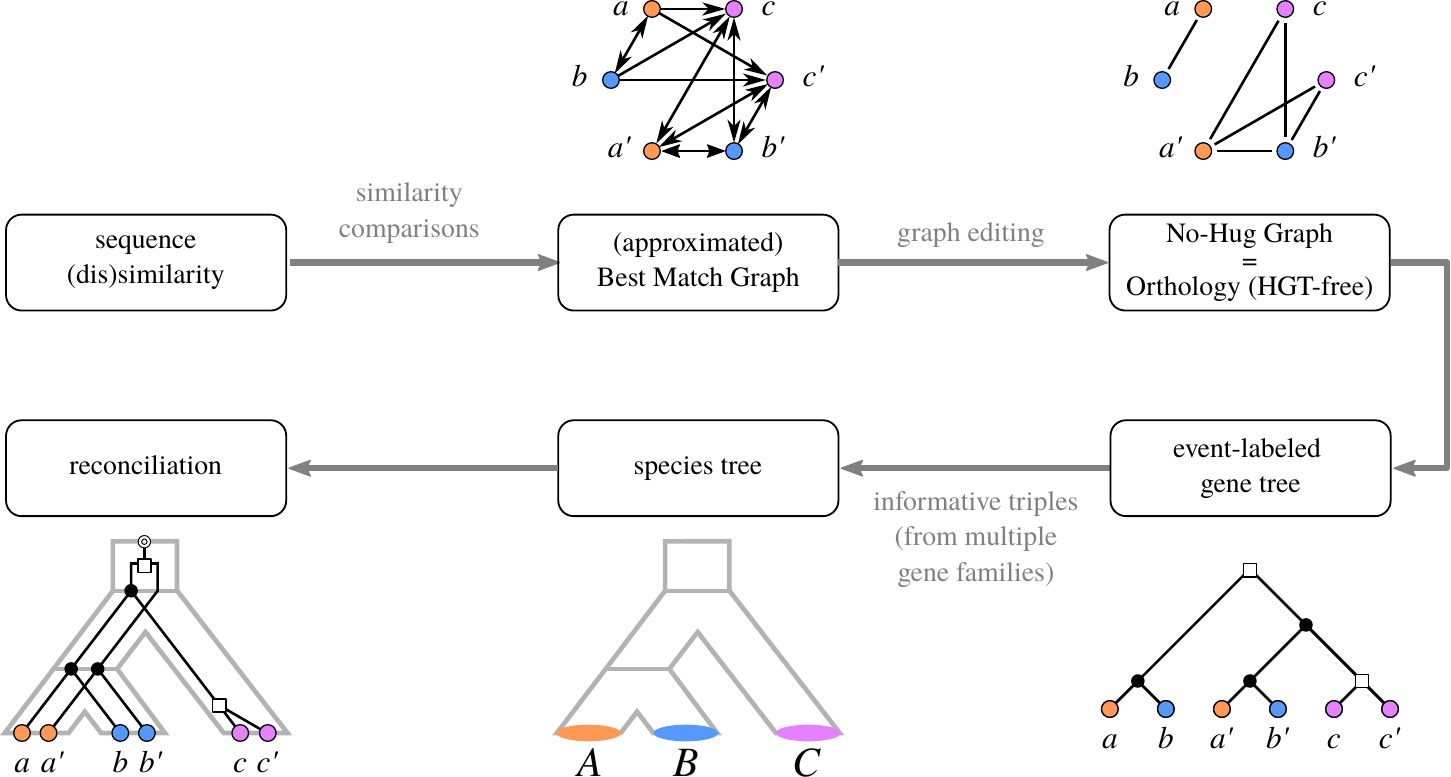}    
  \caption{Conceptual workflow for tree-free orthology detection and
    subsequent reconstruction of event-labeled gene tree $T$, species tree
    $S$, and HGT-free $\sigma$-reconciliation $(T,S,\mu,\sigma)$ under the
    assumption that no HGT occurred.  Here we have $\sigma(x)=\sigma(x')=X$
    for $(x,x',X) \in \{ (a,a',A), (b,b',B), (c,c',C) \}$.  Sequence
    similarity data are used to obtain an initial estimate of the BMG.
    This estimate is corrected to a mathematically sound BMG $(G,\sigma)$
    and the no-hug graph $\mathfrak{N}(G,\sigma)$ is extracted. The no-hug
    graph is free of any false-negative and unambiguously false-positive
    orthology assignments.  In particular, $\mathfrak{N}(G,\sigma)$ is an
    orthology graph and, based on this information, an event-labeled gene
    tree can be reconstructed.  This event labeled gene tree also conveys
    information on the species tree. Integrating the latter information
    over many gene families provides a reliable estimate of the species
    tree. Together with the event labeled gene event labeled gene this
    implies a reconciliation, and thus a complete gene family history.}
  \label{fig:concept-HGTfree}
\end{figure}

\subsection{Clusters of Orthologous Genes} 

Orthologs are often summarized as \emph{clusters of orthologous groups}
(COGs) \cite{Tatusov:97}. We have seen, however, that orthologous genes
form cographs, and hence orthology is in general not a transitive
relation. COGs are only an approximation that is particularly useful
if the the gene family history contains only a small number of
duplications. A special case is one-to-one orthology, where each gene has a
unique ortholog in every other species. In this case, the orthology relation
becomes transitive and the the orthology graph $\Theta$ reduces to a
disjoint union of cliques \cite{Roth:08}. These clusters, e.g.\ computed by
OMA \cite{Roth:08}, are induced complete subgraphs, i.e., cliques in the
full orthology graph. 

Most other approaches to computing COGs allow co-orthologs, i.e., clusters
are not restricted to cliques of the orthology graph and thus, may
include orthologs and paralogs \cite{Tatusov:97}.  A wide variety of
clustering algorithms have been used to extract COGs from sequence
similarity data. The definition of such COGs necessarily depends on
stringency parameters that gauge the trade-off between size and stringency
of COGs. From a theoretical point of view transitivity clustering
\cite{Rahmann:07} is interesting because of its conceptual similarity to
co-graph editing: here the initial orthology estimate is edited by
insertion end deletion of edges to a transitive graph, i.e., a partitioning
into COGs. In \cite{Falls:08}, maximal T{\'u}ran (complete multipartite)
graphs are computed. These form a special class of co-graphs and accommodate
so-called in-paralogs \cite{TremblaySavard:12,Altenhoff:19}, i.e.,
duplicate gene that originated after the most recent speciation event
  in each lineage. Complementarily, it is of interest to partition a gene
set such that out-paralogs (i.e., pairs of genes arising from duplications
that pre-date all speciation events) are placed in different
clusters. These correspond to the connected components of the orthology
graph $\Theta$.

\section{Fitch Graphs and Horizontal Gene Transfer}

\subsection{Definition and Characterization}

Fitch \cite{Fitch:00} defined \emph{``two genes $x$ and $y$ as
\emph{xenologs} if their history, since their common ancestor, involves an
interspecies (horizontal) transfer of the genetic material for at least one
of them.''}  Two leaves in $x,y\in L(T)$ in a tree $T$ are thus xenologs
whenever the unique path connecting $x$ and $y$ in $T$ contains an
HGT-edge. By Def.~\ref{def:HGT}, the subset of HGT-edges $H_{\mu}\subseteq
E(T)$, and thus xenology, depends explicitly on the reconciliation
$(T,S,\mu)$.  As in the previous sections, we are interested in inferring
xenology without first computing a reconciliation $(T,S,\mu)$. Again, we
approach this problem by studying properties of graphs that are implied by
reconciliation or a relaxed scenario.

Let use first consider the parts of a scenario that are unaffected by HGT
events.  Deleting from $T$ all HGT-edges yields a forest and induces a
partition of $L(T)=L_1\cupdot L_2\cupdot \dots\cupdot L_k$ of the leaf set
such that the restriction $T_{|L_i}$ contains no HGT-edges. Consequently,
one can perform the analysis of HGT-free systems outlined above for each of
the gene sets $L_i$ separately. Note that in general, this does not amount
to simply considering the subgraph $(G[L_i],\sigma_{|L_i})$ of empirical
BMGs $(G,\sigma)$ induced by $L_i$. In general the empirical BMGs
$(G_i,\sigma)$ with $V(G_i)=L_i$ will feature best matches that might be
worse than those best matches in $(G[L_i],\sigma_{|L_i})$ that are implied
by $T$ and the additional knowledge of HGT-edges. We shall return to the
inference of HGT-edges and the partitioning of the gene set into maximal
HGT-free subsets below.

Mechanistically HGT is a inherently directional event.  There is a clear
distinction between the horizontally transferred ''copy'' and the
``original'' that continues to be transferred vertically. It is
significant, therefore, whether HGT-edges are found along the path from
$\lca(x,y)$ to $y$, the path from $\lca(x,y)$ to $x$, or along both paths.
Mathematically, this can be captured in the following manner.

\begin{definition} 
  Let $T$ be a rooted tree and $H\subseteq E(T)$ be a set of HGT-edges.
  Then the (directed) Fitch graph $\fitch(T,H)$ has vertex set $L(T)$ and
  $(x,y)\in E(\fitch)$ if the path from $\lca_T(x,y)$ to $y$ contains an
  edge $e\in H$.
\end{definition}
A graph $G$ is a (directed) Fitch graph if $G=\fitch(T,H)$ for some tree
$T$ and edge set $H\subseteq E(T)$. In \cite{Geiss:18} a characterization
of Fitch graphs in terms of eight forbidden subgraphs on three vertices is
given.

Fitch graphs also have a surprisingly simple characterization in terms of
their ``closed complementary neighborhoods''. For a directed graph consider
\begin{equation}
  \overline{N}[x] \coloneqq \{y\in L \mid (x,y)\notin E(G) \},
\end{equation}
Since all graphs considered here are loop-free, we have $x\in\overline
N[x]$. We write $\overline{\mathcal{N}}(G)\coloneqq\{\overline N[y] \mid
y\in L\}$.
\begin{thm}[\cite{HS:19}]
  \label{thm:Seemann}
  $G$ is a Fitch graph if and only if (i) $\overline{\mathcal{N}}(G)$ is
  hierarchy-like, i.e., for $\overline N_1,\overline
  N_2\in\overline{\mathcal{N}}(G)$ hold $\overline N_1\cap \overline
  N_2\in\{\overline N_1,\overline N_2,\emptyset\}$, and (ii) for every
  $\overline N\in\overline{\mathcal{N}}(G)$ and every $y \in\overline N$
  holds $|\overline N[y]| \le |\overline N|$.
\end{thm}
If $G$ is a Fitch graph, then $\mathcal{C}(G)\coloneqq
\overline{\mathcal{N}(G)}\cup \{L\} \cup \{ \{x\}| x\in L\}$, i.e., the
extension of the closed complementary neighborhoods by the singleton sets
and $L$ itself forms a hierarchy on $L$, which corresponds to a uniquely
defined rooted tree $T_G$. This tree $T_G$ is the unique least-resolved
tree for $G$ and the set $H_G$ of HGT-edges in $T_G$ is uniquely determined
and consists of all edges $xy$ in $T_G$ with $x,y\in V^0(T_G)$
\cite{HS:19,Geiss:18}. The directed Fitch graph thus is also informative
about the structure of the gene tree. The pair $(T_G,H_G)$ also immediately
yields the partition of $L$ into HGT-free subsets.

The symmetric version $G_{sym}$ of a directed Fitch graph $G$, called
\emph{undirected Fitch graph}, contains all bi-directional edges $xy\in
E(G)$.  Thus, $xy\in E(G_{sym})$ if and only if the path between $x$ and
$y$ contains a HGT-edge. Undirected Fitch graphs have been studied in
\cite{HLGS:18} and coincide with the complete multipartite graphs. Hence,
undirected Fitch graphs are completely defined by their independent
sets. This implies that undirected Fitch graphs cannot convey much
interesting phylogenetic information because they are completely determined
by the maximal subsets $L_i$ of taxa that have not experienced any
horizontal transfer events among them.

The problem of finding a common gene tree from which a given orthology
graph and Fitch graph can be derived has been considered in
\cite{Hellmuth:21b}.  Moreover, given a leaf-labeled tree $(T,\sigma)$ and
HGT-edges $H\subseteq E(T)$, a result similar in spirit to
Prop. \ref{prop:species-triples-ortho} determines a set triples that must
be displayed the species tree $S$ in any $\sigma$-reconciliation
$(T,S,\mu,\sigma)$ that satisfies \AX{R4}, \AX{R5}, and $H=H_{\mu}$
\cite{Hellmuth:17,Nojgaard:18a}. In \cite{Lafond:20}, a polynomial-time
algorithm is provided that takes $(T,\sigma)$ and $H$ as input and
constructs the species tree $S$ and a time-consistent reconciliation map
$\mu$, if one exists.

\subsection{LDT Graphs} 

Unfortunately, no practical way to infer the Fitch graph directly from
sequence data has become available so far.  A promising approach starts
from the observation that if two genes $x$ and $y$ are more closely related
than the species $\sigma(x)$ and $\sigma(y)$, then $x$ and $y$ must be
xenologs \cite{Novichkov:04}. A statistical sound method for identifying
such pairs of genes is described in \cite{Kanhere:09}.  Although such
indirect methods have been quite successful, see \cite{Ravenhall:15} for a
review, their mathematical properties have been studied only very recently:
\begin{definition}[\cite{Schaller:21f}]\label{def:LDTgraph} 
  Let $(T,\tau_T)$ and $(S,\tau_S)$ be two dates trees and $\sigma:L(T)\to
  L(S)$. Then the undirected graph $\LDT(T,S,\tau_T,\tau_S,\sigma)$ has
  vertex set $L(T)$ and $ab$ is an edge if
  $\tau_T(\lca_T(a,b))<\tau_S(\lca_S(\sigma(a),\sigma(b))$.
\end{definition} 
A vertex-colored graph $(G,\sigma)$ is a \emph{later-divergence-time graph
(LDT graph)} if there is a tuple of dated trees $(T,\tau_T)$ and
$(S,\tau_S)$ and a map $\sigma:L(T)\to L(S)$ such that
$(G,\sigma)=\LDT(T,S,\tau_T,\tau_S,\sigma)$.  LDT graphs have a simple
characterization.  To this end, consider the set $\mathfrak{S}(G,\sigma)$ of
rooted triples $\sigma(x)\sigma(y)|\sigma(z)$ where $\sigma(x)$,
$\sigma(y)$, and $\sigma(z)$ are pairwise distinct, $xy,yz\in E(G)$ and
$xy\notin E(G)$.
\begin{theorem}[\cite{Schaller:21f}]
  A graph $(G,\sigma)$ is an LDT graph if and only if it is properly
  colored cograph and the triple set $\mathfrak{S}(G,\sigma)$ is
  compatible.
\end{theorem}
A polynomial time algorithm to construct a relaxed $\sigma$-scenario for a
given LDT graph is also described in \cite{Schaller:21f}.  Not
surprisingly, the editing problem for LDT-graphs is NP-complete.

An LDT-graph $(G,\sigma)$ is always a subgraph of the undirected Fitch
graph $\fitch_{sym}(T,H_{\mu})$ for any relaxed $\sigma$-scenario
$(T,S,\mu,\tau_T,\tau_S,\sigma)$ that satisfies $(G,\sigma) =
\LDT(T,S,\tau_T,\tau_S,\sigma)$. Thus, LDT graphs cannot contain
false-negative xenologous gene-pairs, and the complement $\overline{G}$ of
an LDT graph $(G,\sigma)$ contains in particular all edges between pairs of
genes that are not separated by a HGT event. In order to apply methods
developed for HGT-free data, we need to find the partition $L=L_1\cupdot
L_2\cupdot \dots\cupdot L_k$ into maximal HGT-free subsets of a given sets
of genes $L$.  In practice, this can be done by using the solution of the
so-called cluster deletion problem applied on $\overline{G}$, i.e., of
deleting a minimum set of edges from $\overline{G}$ such that the resulting
graph $\overline{G}^*$ is a disjoint union of complete graphs
\cite{Shamir:04}.  Defining $L_i$ as the vertex set of the $i$-th clique in
$\overline{G}^*$ then yields the desired partition of $L$ into maximal
HGT-free subsets.  If $G$ (or equivalently $\overline G$) is a cograph,
cluster deletion can be solved in linear time by a greedy algorithm
\cite{Gao:13}. The LDT graph thus can be used to obtain a the partition of
$L(T)$ into HGT-free subsets. Compatibility of such partitions with a given
gene tree as well as inferring the directions of HGT-edges using such
partitions have been the topic in \cite{HSS:22,SHS:23}

\subsection{Orthology in the presence of HGT} 

% orthology relations
\newcommand{\wQO}{\ensuremath{\Psi^{w}}}
\newcommand{\sQO}{\ensuremath{\Psi^{s}}}
\newcommand{\wO}{\ensuremath{\Theta^{w}}}
\newcommand{\sO}{\ensuremath{\Theta^{s}}}

Most of the mathematical results concerning orthology have been obtained in
an HGT-free setting. In the presence of HGT, descendants of two genes that
originate from a speciation event may even eventually reside in the same
species, where they appear as paralogs. This has led to different proposals
for the ``correct'' definition of orthology.  A classification of subtypes
of xenology that, in line with \cite{Fitch:00}, reserves the terms
\emph{ortholog} and \emph{paralog} to situations in which the path between
$x$ and $y$ does not contain an HGT event was proposed in
\cite{Darby:17}. Similar to LDT graphs, ``Equal Divergence Time'' (EDT)
graphs capture that fact that the divergence time of two genes $x$ and $y$
matches the divergence time of the species in which they reside.  Similar
to LDT graphs, a vertex-colored undirected graph $(G,\sigma)$ is an
\emph{EDT graph}, if there are for two dated trees $(T,\tau_T)$ and
$(S,\tau_S)$ and $\sigma:L(T)\to L(S)$ such that $ab \in E(G)$ precisely if
$\tau_T(\lca_T(a,b))=\tau_S(\lca_S(\sigma(a),\sigma(b))$.  In this case, we
also write $(G,\sigma) =\EDT(T,S,\tau_T,\tau_S,\sigma)$.  In general, EDT
graph recognition is NP-hard \cite{SHL+23}, however, it becomes
polynomial-time solvable if information of the LDT graph is available. To
see this, observe that complementary graph of the union of the EDT and LDT
graph contains all edges $ab$ for which
$\tau_T(\lca_T(a,b))>\tau_S(\lca_S(\sigma(a),\sigma(b))$ which determines
the so-called \emph{prior divergence time} graph and we can apply
\cite[Cor.\ 8]{SHL+23}.

One easily verifies that, in a HGT-free $\sigma$-reconciliation
$(T,S,\mu,\sigma)$ that satisfies \AX{R5}, Axiom \AX{R5}(i) implies that
two genes $x$ and $y$ are orthologs if their last common ancestor $\lca_T(x,
y)$ coincides with the last common ancestor $\lca_S(\sigma(x), \sigma(y))$
of the two species in which $x$ and $y$ reside. In this case, there is an
obvious connection between orthology and the EDT graphs. We summarize now
results that shows that EDT graphs are also closely connected to different
notions of orthology in scenarios with HGT that have been discussed in the
literature.  Disagreements on the definitions of orthology in the presence
of HGT stem for the fact that, in general, pairs of genes originating from
a speciation event may be separated by HGT, and thus become xenologs. At
the same time, they may even eventually reside in the same species and
therefore appear as paralogs.

\begin{definition}\label{def:wqo}
  Let $\scen=(T,S,\mu,\tau_T,\tau_S,\sigma)$ be a relaxed
  $\sigma$-scenario. Then, two distinct vertices $x,y\in L(T)$ are
  \begin{itemize}
  \item \emph{weak quasi-orthologs} if $\mu(\lca_T(x,y))\in V^0(S)$.
  \item \emph{weak orthologs} if $\mu(\lca_T(x,y))\in V^0(S)$ and there is
    no HGT-edge along the path between $x$ and $y$ in $T$.
  \item \emph{strict quasi-orthologs} if
    $\mu(\lca_T(x,y))=\lca_S(\sigma(x),\sigma(y))$.
  \item \emph{strict orthologs} if $\mu(\lca_T(x,y)) = \lca_S(\sigma(x),
    \sigma(y))$ and is no HGT-edge along the path between $x$ and $y$ in
    $T$. \smallskip
  \end{itemize}
  The undirected graphs $\wQO(\scen)$, $\wO(\scen)$, $\sQO(\scen)$, and
  $\sO(\scen)$, resp., have vertex set $L(T)$ and edges $xy$ precisely if
  the genes $x$ and $y$ are weak quasi-orthologs, weak orthologs, strict
  quasi-orthologs, resp., strict orthologs in $\scen$, see Table
  \ref{tab:ortho-terminology} for a summary.
\end{definition}

\begin{table}
  \caption{Summary of the alternative notions of orthology in the presence
    of HGT events for a given relaxed $\sigma$-scenario $\scen$.}
  \label{tab:ortho-terminology}
  \centering
  \renewcommand{\arraystretch}{1.3}
  \begin{tabular}{|l||c|c|}
    \hline
    Reconciliation condition & HGT irrelevant  & HGT excluded \\
    \hline
    $\mu(\lca_T(x,y)) \in V^0(S)$  & $\wQO(\scen)$ &  $\wO(\scen)$ \\
    & weak quasi-ortholog  & weak ortholog \\
    \hline
    $\mu(\lca_T(x,y)) = \lca_S(\sigma(x), \sigma(y))$ &
    $\sQO(\scen)$ & $\sO(\scen)$ \\
    & strict quasi-ortholog & (strict) ortholog \\
    \hline
  \end{tabular}
\end{table}

Weak quasi-orthologs are, in essence, Walter Fitch's original, purely
event-based definition of orthology \cite{Fitch:70}. In later work, Fitch
\cite{Fitch:00} emphasizes the condition that ``the common ancestor lies in
the cenancestor (i.e., the most recent common ancestor) of the taxa from
which the two sequences were obtained'', which translates to the notion of
strict quasi-orthologs. Other definitions of orthology explicitly exclude
xenologs \cite{Gray:83,Fitch:00,Darby:17}, which leads the concept of weak
and strict orthologs.

As shown in \cite{SHL+23}, for every  relaxed $\sigma$-scenario $\scen$, we
have
\begin{equation}
  \sO(\scen)\subseteq \sQO(\scen)\subseteq \wQO(\scen) \text{ and }
  \sO(\scen)\subseteq \wO(\scen)\subseteq \wQO(\scen),
\end{equation}
while $\sQO(\scen)$ and $\wO(\scen)$ are incomparable w.r.t.\ the subgraph
relation. Moreover, the weak quasi-orthology graph $\wQO(\scen)$, the weak
orthology graph $\wO(\scen)$ and the strict orthology graph $\sO(\scen)$
are cographs for every relaxed $\scen$. This is not true in general for the
strict quasi-orthology graph $\sQO(\scen)$. There is a close connection
between LDT and EDT graphs and weak and strict orthologs:
\begin{proposition} \cite[Prop.\ 5]{SHL+23}
  Given two graphs $G_1$ and $G_2$ there is a relaxed $\sigma$-scenario
  $\scen$ such that $G_1=\LDT(\scen)$ and $G_2=\EDT(\scen)$ if and only if
  there is relaxed $\sigma$-scenario $\scen'$ with
  $G_1=\LDT(\scen')$ and $G_2=\EDT(\scen')$ that in addition satisfies
  $\wO(\scen')=\sO(\scen')$.   
\end{proposition}
That is, a pair of an LDT and EDT graph can be explained by a common
relaxed scenario, if there the two graphs can be explained a relaxed
scenario for which weak and strict orthology coincide. 

In the generic case, unrelated evolutionary events happen at distinct time
points:
\begin{definition}
  A relaxed scenario is \emph{generic} if $\tau_T(\lca_T(x,y))=\tau_S(u)$
  for $x,y\in L(T)$ and $u\in V^0(S)$, then $\mu(\lca_T(x,y))=u$.
\end{definition}
In generic scenarios, the EDT graph and the different notions of orthology
are connected as follows:
\begin{equation}
  \sO(\scen) \subseteq \sQO(\scen) = \mathrm{EDT} \subseteq \wQO(\scen)
\end{equation}
In the HGT-free case all these graphs coincide if $\mu(\lca_T(x,y))\in
V^0(S)$ implies $\mu(\lca_T(x,y))=\lca_S(\sigma(x),\sigma(y)$.  A full
formal understanding of orthology and its variants in the presence of HGT
is still lacking, however.

\begin{figure}[t]
\noindent\includegraphics[width=\textwidth]{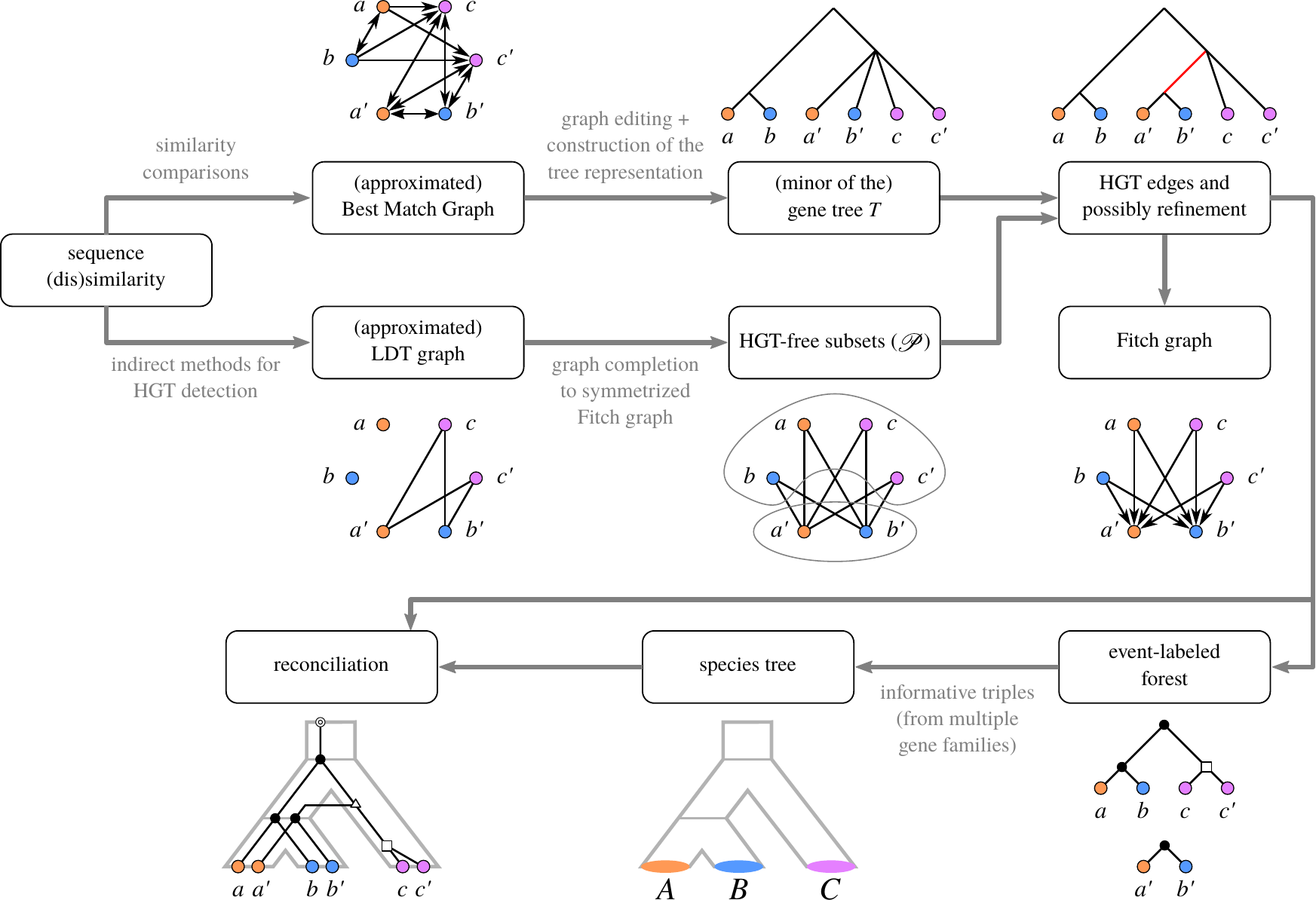}
\caption{Conceptual workflow for tree-free based orthology detection and
  reconstruction of event-labeled gene tree $T$, species tree $S$ and
  $\sigma$-reconciliation $(T,S,\mu,\sigma)$ under the assumption that HGT
  occurred.  Here we have $\sigma(x)=\sigma(x')=X$ for $(x,x',X) \in \{
  (a,a',A), (b,b',B), (c,c',C) \}$.  Sequence similarity data are used to
  obtain an initial estimate of the BMG.  These are corrected to a
  mathematically sound BMG and LDT graph. From these, a usually
  incompletely resolved gene tree and the HGT-free subsets are obtained by
  means of polynomial-time algorithms. In the next step, the workflow as
  outlined in Fig.~\ref{fig:concept-HGTfree} can be applied to the HGT-free
  subsets of genes which results in event-labeled gene trees, which also
  convey information on the species tree. Integrating the latter
  information over many gene families provides a reliable estimate of the
  species tree. Together with the event labeled gene trees this implies a
  reconciliation, and thus a complete gene family history.}
\label{fig:summary}
\end{figure}

\section{Discussion and Open Problems}

In this chapter we provided an overview of the current formal, mathematical
understanding of gene family histories with a focus on ``tree-free''
methods. We deliberately excluded full-fledged probabilistic models from
our discussion. These also start from the idea of scenarios and use
stochastic models of sequence evolution to assign probabilities, which are
then used in maximum likelihood or Bayesian setting to reconstruct
scenarios  \cite{Arvestad:03,Akerborg:09,Larget:10}. While these
detailed models promise accurate results for small and medium data sets, it
is unclear whether they scale to very large problems. The combinatorial
approach outlines here, on the other hand, holds promise to be able to
address gene family histories at global scales.

Taken together, Figs.~\ref{fig:concept-HGTfree} and \ref{fig:summary}
outline a common conceptual workflow for a comprehensive analysis of gene
family histories in a ``tree-free'' setting.  Instead of a gene tree $T$
and a species tree $S$, a BMG and LDT/EDT graphs are estimated from the
data. Together, these two graphs capture the salient information, as
outlined in the formal results in the previous sections. The main advantage
of this approach over the direct reconstruction of trees is that it can
make use of the inherent redundancies in the data. Following the
independent estimate of the edges of the BMG and LDT/EDT graphs, one can
use the fact well-formed BMGs and LDT graphs belong to highly restrictive,
special graph classed, as means of data correction: the initial estimates
can be edited to the closest well-formed graphs of the respective
type. Unfortunately, these graph editing problems are NP-complete. However,
workable heuristics have already been developed for this purpose. The
hardness of this data correction task also does not amount to a conceptual
argument against the workflow proposed in Fig.~\ref{fig:summary}; after
all, both the multiple sequence alignment problem, see \cite{Elias:06} and
the references therein, and the reconstruction of a phylogenetic tree by
means of maximum parsimony or maximum likelihood are also NP-complete
\cite{Day:83,Day:87,Roch:06}. From the BMG and LDT/EDT graphs one can now
obtain a not necessary fully resolved gene tree, sufficient information of
the species, as well as a corresponding event labeling that located all
relevant evolutionary events. In a final step, the resolution of the gene
tree and the species could be improved by conventional phylogenetic methods
using the combinatorially inferred trees $T$ and $S$ as
constraints. Although the time of writing this chapter no full-fledged
implementation have become available, the individual steps have been
benchmarked using simulated data and shown to be feasible in a realistic
setting \cite{Schaller:22a}.

The notion of reconciliation generalizes beyond the scope of this chapter,
which focuses on genes as part of genomes, or, equivalently host-parasite
systems \cite{Page:97,Merkle:10}. The same structure arises when the
evolution of domains in proteins is considered. This naturally gives rise
to a multi-level version of reconciliation \cite{Menet:22,Penel:22}. In
addition to speciation, gene duplication, gene loss, and horizontal
transfer, it is of interest to consider additional event types. Incomplete
lineage sorting or deep coalescence as well as hybridization, in
particular, have received considerable attention
\cite{Stolzer:12,Zheng:14,Chan:17,Du:18,Ansarifar:20}. Interestingly,
maximum parsimony reconciliation becomes NP-hard in this setting
\cite{LeMay:21}. Phylogenetic networks are natural generalization of
phylogenetic trees \cite{HHM:19,huson_rupp_scornavacca_2010}. The concept
of reconciliation persists in such models. In
  \cite{To:15,Scornavacca:17}, for instance, reconciliation of gene trees
  with a restricted classes of species networks are considered. In The
problem of reconciling a gene network (motivated by recombination) with a
species tree is investigated in \cite{Chan:19}, introducing an analog of
the LCA reconciliation that provides a solution for so-called tree-child
networks.  To our knowledge, these extended models have not been studied so
far from a ``tree/network-free'' perspective that would extend the results
reviewed in this chapter.

Despite substantial progress since the first edition of this book
\cite{Setubal:18a}, a number of key questions remain open, in particular in
relation to the detection and localization of HGT events. Most importantly,
the empirically accessible relations (best matches, lower/equal divergence
time) provide independent constraints on the reconciliation $(T,S,\mu)$ and
inferred orthology and xenology relations may impose additional
conditions. In most cases, at least part of the conditions for the
existence of an explaining scenario are expressed in terms of consistence
of triple sets. This suggests consider consistency of their union. In
general this will lead to the NP-complete problem of determining maximum
consistent subset(s) of triple or one of its variants \cite{Byrka:10}.  A
better understanding how conflicting constraints arise, however, may point
to a more accurate and possibly computationally more efficient way of
handling internal conflicts in the data. Complementarily, it will be of
interest to investigate how a trusted species tree could be used for the
improvement of initial estimated of best match graphs. The fact that
information of this type can be helpful has been demonstrated in a
maximum-likelihood framework \cite{Morel:20}. An interesting variation on
this theme are methods to infer species trees and reconciliation maps from
a set of gene trees \cite{Morel:22}. It also remains unclear whether for
given $T$, $S$, and $\sigma$, there exists a reconciliation with a
prescribed set of reconciliation edges $H\subseteq E(T)$. We conjecture
that, in analogy with the HGT-free case, the answer is affirmative. More
generally, event labelings and their consequences are not full understood
for relaxed scenarios. While event labeling are implicitly defined as part
of reconciliation maps e.g.\ in \cite{Tofigh:11,Bansal:12,Stolzer:12},
these definitions also restrict the scenarios to which they pertain, see
also \cite{Nojgaard:18a,Lafond:20}.

From a practical point of view, the most pressing issue is the development
of more efficient heuristics for the graph-editing problems that naturally
arise in any real-life application of the tree-free methods. i.e., in the
intermediate step of the workflows outlined in
Figs.~\ref{fig:concept-HGTfree} and \ref{fig:summary}. For both cographs
\cite{White:18,Hellmuth:20b,Crespelle:21} and BMGs \cite{Schaller:21g}
signficant progress has been reported in the recent past. Full-fledged
bioinformatics pipelines for large-scale applications focus primarily on
COGs rather than a more fine grained presentation of gene family histories,
making only limited use of the progress in mathematical phylogenetics. In
conjunction, with much improved methods for genome-wide alignments
\cite{Armstrong:20}, improved gene family histories may constitute also an
important step towards a comprehensive understanding of genome evolution.
  
\bibliographystyle{unsrt}
\bibliography{orthonew}

\newpage
\printindex
\newpage

\end{document}